\documentclass[sigconf, nonacm]{acmart}

\usepackage{graphicx}
\usepackage{subfigure}
\usepackage{amsthm}
\usepackage[ruled,linesnumbered,vlined,noend]{algorithm2e}
\usepackage{color}
\usepackage{tikz}
\usepackage{tikz-qtree}
\usepackage{pgfplots}
\usepackage{pgfplotstable}
\usetikzlibrary{patterns}
\usepackage{tabularx}
\usepackage{makecell}
\usepackage{multirow}
\usepackage{booktabs}
\usepackage{bm}
\usepackage{balance}
\usepackage{stfloats}
\PassOptionsToPackage{table,xcdraw}{xcolor}
\usepackage{caption} 
\usepackage{enumitem}

\pgfplotsset{compat=1.18}

\usepackage{pifont}       
\usepackage{utfsym} 

\makeatletter
\def\@ACM@checkaffil{
    \if@ACM@instpresent\else
    \ClassWarningNoLine{\@classname}{No institution present for an affiliation}%
    \fi
    \if@ACM@citypresent\else
    \ClassWarningNoLine{\@classname}{No city present for an affiliation}%
    \fi
    \if@ACM@countrypresent\else
        \ClassWarningNoLine{\@classname}{No country present for an affiliation}%
    \fi
}
\makeatother

\definecolor{linecolor}{rgb}{0.55, 0, 0} 
\definecolor{bgcolor}{rgb}{1, 0.95, 0.95} 
\newcommand{\Yes}{\usym{2713}}
\newcommand{\No}{\usym{2717}}

\usepackage{listings}

\usepackage[most]{tcolorbox}
\tcbuselibrary{listings, breakable, skins}

\lstnewenvironment{PromptBox}
{
    \lstset{
        basicstyle=\ttfamily\small, 
        backgroundcolor=\color{white},     
        frame=single,                      
        rulecolor=\color{black},           
        framerule=0.8pt,                   
        breaklines=true,                   
        breakatwhitespace=true,            
        columns=fullflexible,              
        showstringspaces=false,            
        escapeinside={(*@}{@*)},           
        xleftmargin=0.5em,                 
        xrightmargin=0.5em,                
        aboveskip=1em,                     
        belowskip=1em                      
    }
}{}

\tcbset{highlightcode/.style=
    {enhanced, 
    colframe=gray!20, 
    colback=gray!20, 
    boxrule=0pt, 
    boxsep=0pt, 
    left=1mm, 
    right=0mm, 
    top=1mm, 
    bottom=1mm,
    overlay={
        \begin{scope}[shift={(frame.north west)}]
            \foreach \i in {1,2,3} {
                \node[anchor=east] at (-5mm,{-\i*15mm + 1.5mm}) {\footnotesize\the\numexpr\value{AlgoLine}+\i\relax};
            }
        \end{scope}
    },
}}

\usepackage{float}
\usepackage{xspace}
\tcbset{
  aibox/.style={
    width=474.18663pt,
    top=10pt,
    colback=white,
    colframe=black,
    colbacktitle=black,
    enhanced,
    center,
    attach boxed title to top left={yshift=-0.1in,xshift=0.15in},
    boxed title style={boxrule=0pt,colframe=white,},
  }
}
\newtcolorbox{AIbox}[2][]{aibox,title=#2,#1}

\newcommand{\modelname}{BookRAG}

\tcbset{
  ragstyle/.style={
    colback=white,
    colframe=black,
    boxrule=0.8pt,
    toprule=1.5pt,
    bottomrule=1.5pt,
    arc=0mm,
    left=2mm,
    right=2mm,
    top=2mm,
    bottom=2mm,
    width=\textwidth,
    enlarge left by=0mm,
    before skip=1em,
    after skip=1em
  }
}

\newcommand{\caseheader}[1]{%
  \vspace{0.3em}%
  \noindent\colorbox{gray!15}{%
    \parbox{\dimexpr\linewidth-2\fboxsep\relax}{%
      \centering\textbf{#1}%
    }%
  }%
  \vspace{0.3em}%
}

\newcommand{\intrasep}{%
  \par\vspace{0.3em}%
  \noindent\tikz{\draw[dashed, gray!80, line width=0.8pt] (0,0) -- (\linewidth,0);}%
  \par\vspace{0.3em}%
}

\definecolor{mycolor2}{RGB}{164,224,187} 
\definecolor{mycolor1}{RGB}{53,122,162} 

\newcommand{\colorcell}[1]{%
    \cellcolor{mycolor1!#1!mycolor2}%
    \ifnum#1>49%
        {\huge \textcolor{white}{{#1}}}
    \else
        {\huge \textcolor{black}{{#1}}}
    \fi
}

\begin{document}

\title{BookRAG: A Hierarchical Structure-aware Index-based Approach for Retrieval-Augmented Generation on Complex Documents}


\author{Shu Wang}
\affiliation{%
    \institution{The Chinese University of Hong Kong, Shenzhen}
}
\email{shuwang3@link.cuhk.edu.cn}

\author{Yingli Zhou}
\affiliation{%
  \institution{The Chinese University of Hong Kong, Shenzhen}
}
\email{yinglizhou@link.cuhk.edu.cn}

\author{Yixiang Fang}
\affiliation{%
  \institution{The Chinese University of Hong Kong, Shenzhen}
}
\email{fangyixiang@cuhk.edu.cn}

\newcommand{\zhou}[1]{\textcolor{magenta}{[zhou: #1]}}



\begin{abstract}
As an effective method to boost the performance of Large Language Models (LLMs) on the question answering (QA) task, Retrieval-Augmented Generation (RAG), which queries highly relevant information from external complex documents, has attracted tremendous attention from both industry and academia.
Existing RAG approaches often focus on general documents, and they overlook the fact that many real-world documents (such as books, booklets, handbooks, etc.) have a hierarchical structure, which organizes their content from different granularity levels, leading to poor performance for the QA task.
To address these limitations, we introduce {\modelname}, a novel RAG approach targeted for documents with a hierarchical structure, which exploits logical hierarchies and traces entity relations to query the highly relevant information.
Specifically, we build a novel index structure, called BookIndex, by extracting a hierarchical tree from the document, which serves as the role of its table of contents, using a graph to capture the intricate relationships between entities, and mapping entities to tree nodes.
Leveraging the BookIndex, we then propose an agent-based query method inspired by the Information Foraging Theory, which dynamically classifies queries and employs a tailored retrieval workflow.
Extensive experiments on three widely adopted benchmarks demonstrate that {\modelname} achieves state-of-the-art performance, significantly outperforming baselines in both retrieval recall and QA accuracy while maintaining competitive efficiency.
\end{abstract}

\maketitle



\section{Introduction}

\begin{figure}[t]
    \centering
    \includegraphics[width=\linewidth]{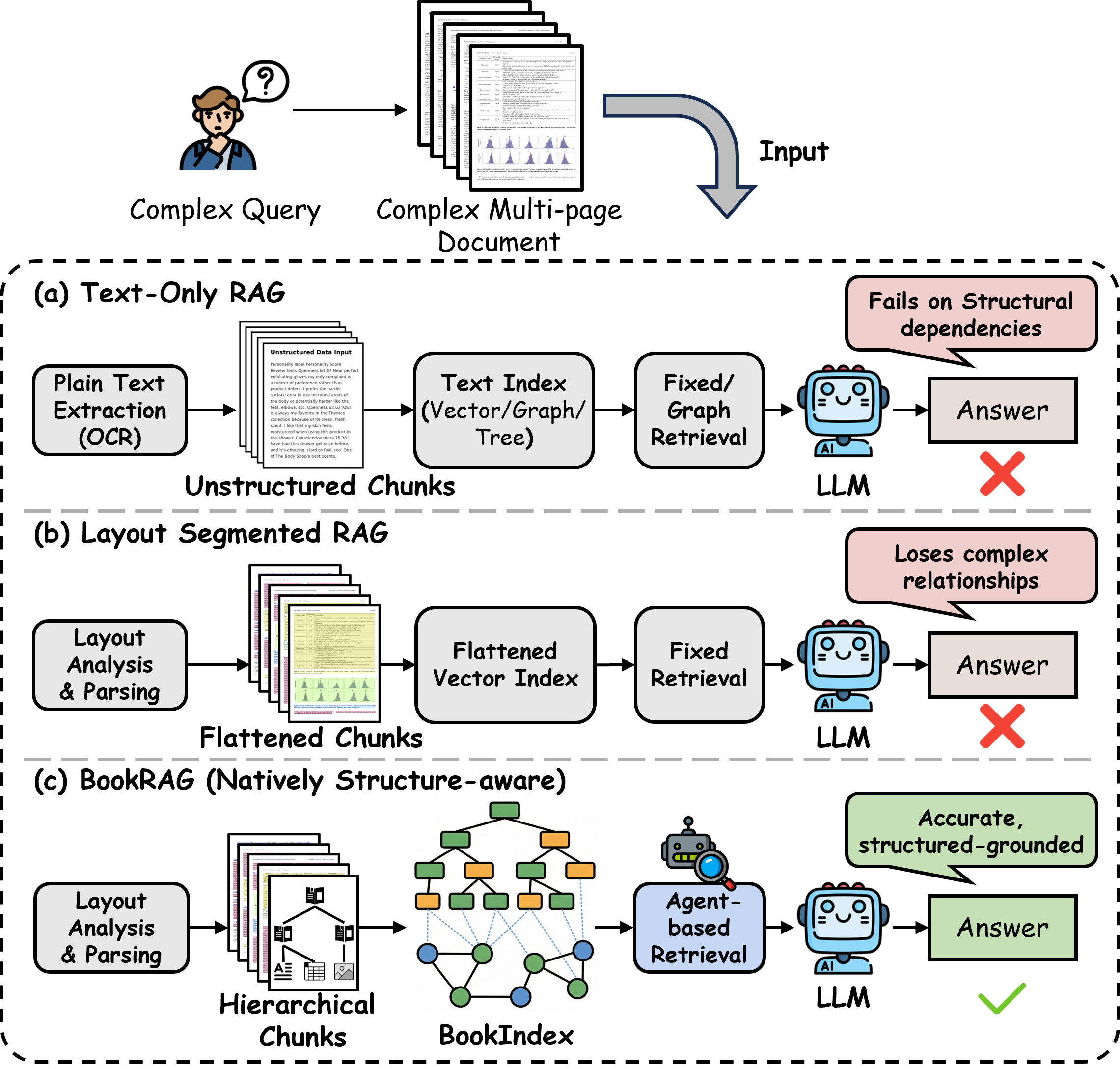}
    \caption{Comparison of existing methods and {\modelname} for complex document QA.}
    \label{fig:alpha}
\end{figure}


Large Language Models (LLMs) such as Qwen 3~\cite{yang2025qwen3} and Gemini 2.5~\cite{comanici2025gemini} have revolutionized the Question Answering (QA) system \cite{daull2023complex,yue2025survey,zhao2023survey}.
The industry has increasingly adopted LLMs to build QA systems that assist users and reduce manual effort in many applications~\cite{zhao2023survey,zhu2023large}, such as financial auditing~\cite{nie2024survey,li2023large}, legal compliance~\cite{chalkidis2020legal}, and scientific discovery~\cite{wang2024large}. 
However, directly relying on LLMs may lead to missing domain knowledge and generating outdated or unsupported information. To address these issues, Retrieval-Augmented Generation (RAG) has been widely adopted~\cite{hu2024rag,gao2023retrieval} by retrieving relevant domain knowledge from external sources and using it to guide the LLM during response generation. 
On the other hand, in real-world enterprise scenarios, domain knowledge is often stored in long-form documents, such as technical handbooks, API reference manuals, and operational guidebooks~\cite{SolutionsReview2019}.
A notable feature of such documents is that they follow the structure of books, characterized by intricate layouts and rigorous logical hierarchies (e.g., explicit tables of contents, nested chapters, and multi-level sections).
In this paper, we aim to design an effective RAG system for QA over long and highly structured documents.

$\bullet$ {\bf Prior works.} The existing RAG approaches for document-level QA generally fall into two paradigms, as illustrated in Figure~\ref{fig:alpha}.
The first paradigm relies on OCR (Optical Character Recognition) to convert the document into plain text, after which any text-based RAG method can be directly applied. 
Among text-based RAG methods, state-of-the-art approaches increasingly adopt \textit{graph-based RAG}~\cite{zhou2025depth, 10.14778/3748191.3748194,zhang2025survey}, where graph data serves as an external knowledge source because it captures rich semantic information and the relational structure between entities. As shown in Table~\ref{tab:rag_comparison}, two representative methods are GraphRAG~\cite{edge2024local} and RAPTOR~\cite{sarthi2024raptor}.
Specifically, GraphRAG first constructs a knowledge graph (KG) from the textual corpus, and then applies the Leiden community detection algorithm~\cite{traag2019louvain} to obtain hierarchical clusters. Summaries are generated for each community, providing a comprehensive, global overview of the entire corpus. 
RAPTOR builds a recursive tree structure by iteratively clustering document chunks and summarizing them at each level, enabling the model to capture both fine-grained and high-level semantic information across the corpus.

In contrast, the second paradigm, \textit{layout-aware segmentation}~\cite{barboule2025survey,wang2024mineru}, first parses the document into structured blocks that preserve the original layout and information of the document, such as paragraphs, tables, figures, or equations.
By doing so, it not only avoids the fixed chunk size used in the first paradigm, which often leads to fragmented information, but also retains document-native structural information.
These blocks often exhibit multimodal characteristics, and a typical approach is to apply multimodal retrieval to obtain relevant content for answering queries. Recently, a state-of-the-art method in this category, DocETL~\cite{shankar2024docetl}, provides a declarative interface that allows users to manually define LLM-based processing pipelines to analyze the retrieved blocks. 
These pipelines consist of LLM-powered operations combined with task-specific optimizations.

%

%
%




%

\begin{table*}[t]
    \centering
    \setlength{\tabcolsep}{6pt} 
    \caption{Comparison of representative methods and our {\modelname}.}
    \begin{tabular}{llcccc}
        \toprule
        \textbf{Type} & \textbf{\makecell[l]{Representative\\ Method}} & \textbf{Core Feature} & \textbf{\makecell[l]{Multi-hop \\ Reasoning}} & \textbf{\makecell{Document\\ Parsing}} & \textbf{\makecell{Query \\ Workflow}} \\
        \midrule
        \multirow{2}{*}{Graph-based} 
        & RAPTOR~\cite{sarthi2024raptor} & Recursive summarization & \Yes & \No & Static \\
        & GraphRAG~\cite{edge2024local} & Global community detection & \Yes & \No & Static \\
        \midrule
        \multirow{2}{*}{Layout segmented} 
        & MM-Vanilla & Multi-modal retrieval & \No & \Yes & Static \\
        & DocETL~\cite{shankar2024docetl} & LLM-based document processing pipeline & \No & \Yes & Manual \\
        \midrule
        \textbf{Doc-Native} & \textbf{{\modelname} (Ours)} & \makecell{Structure-award Index \& Agent-based retrieval} & \Yes & \Yes & Dynamic \\
        \bottomrule
    \end{tabular}
    \label{tab:rag_comparison}
\end{table*}

%

%

%

$\bullet$ {\bf Limitations of existing works.} However, these methods suffer from two fundamental limitations (\textbf{L} for short):
\textbf{L1: Failure to capture the deep connection of document structure and semantics.}
Text-based approaches cannot capture the structural layout of the document, resulting in the loss of important relationships stored in the hierarchical blocks, such as tables nested within a specific section.
While layout-segmented methods preserve document structure, they cannot capture the relationships between different blocks in the document, which limits their capability for multi-hop reasoning across these blocks and ultimately affects their overall performance.
 \textbf{L2: Static of query workflows.}
In real-world QA scenarios, user queries are highly heterogeneous, ranging from simple keyword lookups to complex multi-hop questions that require synthesizing evidence scattered across different parts of the document. Applying a uniform strategy, such as static or manually predefined workflows, to diverse needs is inefficient; for example, complex queries often require question decomposition, whereas simple queries do not.

$\bullet$ {\bf Our technical contributions.} 
To bridge this gap, we introduce \textbf{\modelname}, the first retrieval-augmented generation method built upon a document-native \textbf{BookIndex}, designed to document QA tasks.
Specifically, to capture the deep connection of the relation in the document, BookIndex organizes information through two complementary structures.
First, to preserve the document's native logical hierarchy, we organize the parsed content blocks into a hierarchical tree structure, which serves as the role of its table of contents.
Second, to capture the intricate relations within these blocks, we construct a KG containing fine-grained entities.
Finally, we unify these two structures by mapping the KG entities to their corresponding tree nodes.

However, effective multi-hop reasoning on the graph relies on a high-quality KG~\cite{zhou2025depth,zhang2025survey}, which is often compromised by entity ambiguity (e.g., distinct entities with names like ``LLM'' and ``Large Language Model'').
To address this, we propose a novel gradient-based entity resolution method that analyzes the similarity distribution of candidate entities.
By identifying sharp drops in similarity scores, we can efficiently distinguish and merge coreferent entities, thereby ensuring graph connectivity and enhancing reasoning capabilities.

Building upon the BookIndex, we address the static of query workflows (\textbf{L2}) by implementing an \textbf{agent-based retrieval}.
Specifically, our agent first classifies user queries based on their intent and complexity, and then dynamically generates tailored retrieval workflows.
Grounded in \textit{Information Foraging Theory}~\cite{pirolli1995information}, our retrieval process mimics foraging by using \textit{Selector} to narrow down the search space via information scents and \textit{Reasoner} to locate highly relevant evidence.

We conduct extensive experiments on three widely adopted datasets to validate the effectiveness and efficiency of our {\modelname}, comparing it against several state-of-the-art baselines.
The experimental results demonstrate that {\modelname} consistently achieves superior performance in both retrieval recall and QA accuracy across all datasets.
Furthermore, our detailed analysis validates the critical contributions of our key features, such as the high-quality KG and the agent-based retrieval mechanism.
%

We summarize our contributions as:
\begin{itemize}
    \item We introduce \textbf{\modelname}, a novel method that constructs a document-native \textit{BookIndex} by integrating a hierarchical tree of document layout blocks with a KG storing fine-grained entity relations.
    
    \item We propose an \textbf{Agent-based Retrieval} approach inspired by Information Foraging Theory, which dynamically classifies queries and configures optimal retrieval workflows to locate highly relevant evidence within documents.
    
    \item Extensive experiments on multiple benchmarks show that \textit{\modelname} significantly outperforms existing baselines, attaining state-of-the-art performance in solving complex document QA tasks while maintaining competitive efficiency.
\end{itemize}

\textbf{Outline.}
%
We review related work in Section~\ref{sec:related-work}.
Section~\ref{sec:preliminary} introduces the problem formulation, IFT, and RAG workflow.
In Section~\ref{sec:index}, we present the structure of our BookIndex and its construction.
Section~\ref{sec:bookrag} presents our agent-based retrieval, elaborating on the query classification and operators used in the structured execution of {\modelname}.
We present the experimental results and detailed analysis in Section~\ref{sec:exp}, and conclude the paper in Section~\ref{sec:conclusion}.

\label{sec:background}

\section{Related Work}
\label{sec:related-work}

In this section, we review the related works, including LLM in document analysis and the modern representative RAG approaches.

$\bullet$ \textbf{LLM in document analysis.} Recent advances in LLMs have offered opportunities to leverage LLMs in document data analysis. Due to the robust semantic reasoning capabilities of LLMs, there is an increasing number of works focusing on transferring unstructured documents (e.g., HTML, PDFs, and raw text) into structured formats, such as relational tables \cite{arora2023language,jin2025elt,nobari2024tabulax,chai2025doctopus}.  For example, Evaporate \cite{arora2023language} utilizes LLMs to synthesize extraction code, enabling cost-effective conversion of semi-structured web documents into structured databases without heavy manual annotation.
In addition, several LLM-based document analysis systems have been proposed to equip standard data pipelines with semantic understanding \cite{patel2025semantic,shankar2024docetl,wang2025aop,li2025data+}. For instance, LOTUS \cite{patel2025semantic} extends the relational model with semantic operators, allowing users to execute SQL-like queries with LLM-powered predicates (e.g., filter, join) over unstructured text corpora. Similarly, DocETL \cite{shankar2024docetl} introduces an agentic framework to optimize complex information extraction tasks.
Furthermore, another line of research proposes to directly analyze or parse documents by viewing the document pages as images, thereby preserving critical layout and visual information \cite{kim2022ocr,deepseekvl,wang2024qwen2}.

$\bullet$ \textbf{RAG approaches.} RAG has been proven
to excel in many tasks, including open-ended question answering \cite{jeong2024adaptive,siriwardhana2023improving}, programming context \cite{chen2024auto,chen2023haipipe}, SQL rewrite~\cite{lillm,sun2024r}, and data cleaning~\cite{naeem2024retclean,narayan2022can,qian2024unidm}.
The naive RAG technique relies on retrieving query-relevant contexts from external knowledge bases to mitigate the ``hallucination'' of LLMs.
Recently, many RAG approaches~\cite{edge2024local,guo2024lightrag,gutierrez2024hipporag,sarthi2024raptor,wang2024knowledge,he2024g,ma2024think,ma2024think,li2024dalk,zhou2025depth,wang2025archrag} have adopted graph structures to organize the information and relationships within documents, achieving improved overall retrieval performance.
For more details, please refer to the recent survey of graph-based RAG methods~\cite{peng2024graph}.
Besides, the Agentic RAG paradigm has been widely studied, employing autonomous agents to dynamically orchestrate and refine the RAG pipeline, thus significantly boosting the reasoning robustness and generation fidelity \cite{selfrag,crag,adaptiverag}.

\section{Preliminaries}
\label{sec:preliminary}

This section formalizes the research problem of complex document QA, introduces the foundational Information Foraging Theory (IFT), and briefly reviews the general workflow of RAG systems

\subsection{Problem Formulation}

We study the problem of Question Answering (QA) over complex documents, which aims to answer user queries based on long-form documents~\cite{ma2024mmlongbench,barboule2025survey,cho2024m3docrag}.
Formally, a document $D$ is represented as a sequence of $N$ pages, $D = \{P_i\}_{i=1}^N$.
These pages collectively contain a sequence of content blocks $\mathcal{B} = \{b_j\}_{j=1}^M$, where each block $b_j$ represents a distinct element (e.g., text segment, section header, table, or image) organized within a logical chapter hierarchy.
Given a user query $q$, the goal is to generate an accurate answer $A$, ideally grounded in a specific set of evidence blocks $E \subset \mathcal{B}$.
The task is formulated as developing a method $\mathcal{S}$ that maps the structured document and the query to the final answer:
\begin{equation}
    A = \mathcal{S}(D, q)
\end{equation}
where $\mathcal{S}$ should navigate both the sequential page content and the logical hierarchy of $D$ to synthesize the response.

\begin{figure*}
    \centering
    \includegraphics[width=1\linewidth]{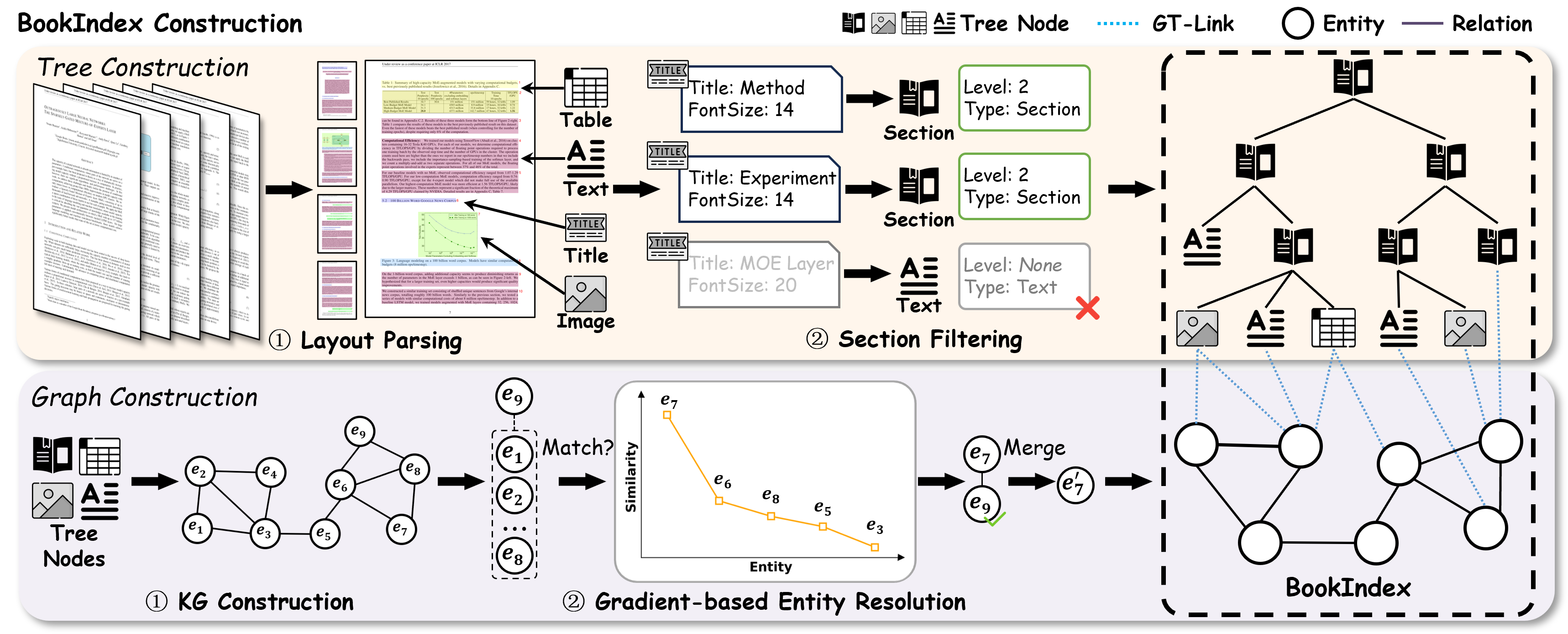}
    \caption{The BookIndex Construction process. This phase includes Tree Construction, derived from Layout Parsing and Section Filtering, and Graph Construction, which involves KG Construction and Gradient-based Entity Resolution.}
    \label{fig:overview}
\end{figure*}

\subsection{Information Foraging Theory}

Information Foraging Theory (IFT)~\cite{pirolli1995information} provides a framework for understanding information access as a process analogous to animal foraging.
It suggests that users follow cues, known as \textbf{information scent} (e.g., keywords or icons), to navigate between clusters of content, known as \textbf{information patches} (e.g., sections in handbooks). 
The goal is to maximize the rate of valuable information gain while minimizing effort, guiding the decision to either stay within a patch or seek a new one.

Consider experts seeking a solution to a specific problem within a large technical handbook. They first extract key terms related to the problem, which act as information scent. 
This scent guides them to navigate towards one or more promising sections (the information patches). 
Within these patches, they analyze the diverse content to extract the precise knowledge required to formulate a final answer



\subsection{RAG workflow}

Retrieval-Augmented Generation (RAG) systems typically operate in a two-phase framework~\cite{peng2024graph,edge2024local,10.14778/3748191.3748194}.
In the \textbf{Offline Indexing} phase, unstructured corpus data is organized into a structured index, which can take various forms such as vector databases or KG~\cite{zhou2025depth}.
Subsequently, in the \textbf{Online Retrieval} phase, the system retrieves relevant components (e.g., text chunks or subgraphs) based on the user query $q$ to inform the LLM's generation.
However, these general workflows often treat the index as a structure derived purely from content, potentially detaching it from the document's original logical hierarchy.
In contrast, our approach seeks to deeply integrate these retrieval structures with the document's native tree topology.



\section{BookIndex}
\label{sec:index}

This section introduces our proposed \textbf{BookIndex}, a hierarchical structure-aware index designed to capture both the explicit logical hierarchy and the intricate entity relations within complex documents.
We first formally define the structure of the BookIndex ($B$).
Subsequently, we elaborate on the sequential, two-stage construction process:
(1) \textbf{Tree Construction}, which parses the document's layout to establish a hierarchical nodes, each categorized by type; and
(2) \textbf{Graph Construction}, which extracts fine-grained entity knowledge from the tree nodes and refines it through a novel gradient-based entity resolution method.
%

\subsection{Overview of BookIndex}

We formally define our \textbf{BookIndex} as a triplet $B = (T, G, M)$.
Here, $T = (N, E_T)$ represents a \textit{Tree} structure where $N$ is the set of nodes derived from the document's explicit logical hierarchy (e.g., titles, sections, tables), and $E_T$ denotes their nesting relationships.
$G = (V, E_G)$ is a \textit{Knowledge Graph} that captures fine-grained entities ($V$) and their relations ($E_G$) scattered throughout the document.
Finally, $M: V \to \mathcal{P}(N)$ is the \textit{Graph-Tree Link (GT-Link)}, which links each entity in $V$ to the set of specific tree nodes in $N$ from which it was extracted.
These links are crucial for capturing the intricate, cross-sectional relations within the document.
The hierarchical tree nodes in $T$ serve as the document's native \textit{information patches}, providing structured contexts for information seeking.
Meanwhile, the entities and relations in $G$, connected via $M$, act as the rich \textit{information scent} that guides navigation between and within these patches.

Figure~\ref{fig:overview} provides an example of our BookIndex.
The Tree component, positioned at the top, organizes the document into a hierarchical structure, where content blocks such as text, tables, and images serve as leaf nodes nested within section nodes.
The Graph component is composed of entities and relations extracted from these nodes.
The GT-Link, illustrated by the blue dotted lines, explicitly connects these entities back to their corresponding tree nodes, thereby grounding the semantic entities within the document's logical hierarchy.

\subsection{Tree Construction}

The first stage transforms the raw document into a structured hierarchical tree $T$. This involves two key steps: robust layout parsing and intelligent section filtering.

\subsubsection{Layout Parsing}
The Layout Parsing phase processes the input document $D$ (a collection of pages) using layout analysis and recognition models.
This step identifies, extracts, and organizes diverse blocks (e.g., text, tables, images) from the document pages.
The output is a sequence of primitive blocks, $\mathcal{B} = \{b_1, b_2, \cdots, b_k\}$, where each block $b_i = (c_i, \tau_i, f_i)$ is defined as a triplet.
Here, $c_i$ is the raw content (e.g., text, image data), $\tau_i$ is the initial layout-based type (e.g., \texttt{Title, Text, Table, Image}), and $f_i$ is a vector of associated layout features (e.g., ``FontSize'', bounding box).

\subsubsection{Section Filtering.}
Next, the {Section Filtering} phase processes this initial sequence to identify the document's logically hierarchical structure.
Layout Parsing identifies blocks as \texttt{Title} but does not assign their hierarchical level.
Therefore, we select the candidate subset $\mathcal{B}_{\text{title}} \subset \mathcal{B}$ (where $\tau_i = \texttt{Title}$) for an LLM-based analysis.
To handle extremely long documents, this analysis is performed in batches, where each batch retains a contextual window of high-level section information (with $l=1$ as the root).
The LLM analyzes the content $c_i$ and layout features $f_i$ of the candidates to determine two key properties: their actual hierarchical level $l_i \in \{1, 2, ...\}$ and final node type $\tau'_i$ (e.g., re-classifying an erroneous \texttt{Title} as \texttt{Text} if its level is ``None'').
This step is crucial for preserving the document's logical hierarchy by correcting blocks erroneously parsed as \texttt{Title}, such as descriptive text within images or borderless table headers.

Finally, the definitive tree $T = (N, E_T)$ is constructed.
The node set $N$ is composed of all blocks from the filtering and re-classification process, where each node $n \in N$ retains its content ($c_i$) and its final node type ($\tau'_i$) (e.g., \texttt{Text}, \texttt{Section}, \texttt{Table}, and \texttt{Image}). 
The edge set $E_T$, representing the parent-child nesting relationships, is then established. 
Parent-child relationships are inferred by sequentially traversing the nodes, using both the determined hierarchical levels ($l_i$) of \texttt{Section} nodes and the overall document order to assemble the complete tree structure.

As an example shown in Figure~\ref{fig:overview}, the \textit{Layout Parsing} phase identifies diverse blocks, typing them as \texttt{Title}, \texttt{Text}, \texttt{Table}, and \texttt{Image}. 
During the \textit{Section Filtering} phase, the \texttt{Title} candidates (e.g., "Method", "Experiment", and "MOE Layer") are analyzed by the LLM.
The blocks ``Method'' and ``Experiment'' (both with ``FontSize: 14'') are correctly identified as \texttt{Section} nodes at ``Level: 2''.
Conversely, the ``MOE Layer'' block (``FontSize: 20''), which was erroneously tagged as \texttt{Title} by the parser, is re-classified by the LLM as a \texttt{Text} node with ``Level: None''.
This correction is crucial for preserving the document's logical hierarchy.
Following this process, all filtered and classified nodes are assembled into the final tree structure based on their determined levels and document order.

\subsection{Graph Construction}

Once the tree $T$ is established, we proceed to populate the Knowledge Graph $G$ by extracting and refining entities from the tree nodes.

\subsubsection{KG Construction.}
We iterate each node $n_i \in N$ from the previously constructed tree $T$.
For each node $n_i$, we extract a subgraph $g_i = (V_i, E_{Ri})$ based on its content $c_i$ and final node type $\tau'_i$.
This extraction is modality-dependent: if the node is text-only, an LLM is prompted to extract entities and relations, while for nodes containing visual elements (e.g., $\tau'_i = \texttt{Image}$), a Vision Language Model (VLM) is employed to extract visual knowledge.
Crucially, for every entity $v \in V_i$ extracted, its origin tree node $n_i$ is recorded, which is vital for constructing the final mapping $M$.

Furthermore, to preserve structural semantics for specific logical types (e.g., \texttt{Table}, \texttt{Formula}), our process first creates a distinct, typed entity (e.g., $v_{\text{table}}$ representing the table itself).
The other extracted entities from the specific node's content are linked to this primary vertex.
For \texttt{Table} nodes specifically, row and column headers are also explicitly extracted as distinct entities and linked to $v_{\text{table}}$ via a ``ContainedIn'' relationship.

\subsubsection{Gradient-based Entity Resolution.}
As shown in the literature~\cite{zhou2025depth,zhang2025survey}, a well-constructed KG is essential for document question answering.
A common challenge in the extraction process is that the same conceptual entity is often fragmented into multiple distinct entities due to abbreviations, co-references, or its varied occurrences across different document sections.
This necessitates a robust Entity Resolution (ER) process, which identifies and merges these fragmented entities to refine the raw KG.

\begin{algorithm}[t]
  \caption{Gradient-based entity resolution}
  \label{alg:er}
  \small
    \KwIn{KG ${G}$, 
    New entity $v_n$,
    Rerank model $\mathcal{R}$, Entity vector database $DB$, Vector search number $top\_k$, threshold of gradient $g$}
    \tcp{\textcolor{teal}{Vector Search $top\_k$ relevant entities in $DB$.}}
    $E_c \gets$ Search($DB, v_n, top\_k$)\;
    
    $\mathcal{S}\gets\mathcal{R}(E_c, v_n)$\;
    \tcp{\textcolor{teal}{Sort all candidate entities by rerank scores.}}
    Sort($E_c, \mathcal{S}$)\;
    $score\gets\mathcal{S}[0]$, $Sel \gets E_c[0]$\;
    
    \tcp{\textcolor{teal}{Gradient select similar entities.}}
    \For{each remain entity $v_c \in E_c\setminus\{E_c[0]\}$}{
        \If{$\mathcal{S}[v_c]>score/g$}{
            $Sel\gets Sel \cup \{v_c\}$, $score\gets \mathcal{S}[v_c]$\;
        }
        \lElse{break}
    }
    
    \tcp{\textcolor{teal}{Merge entity or add new entity.}}
    \If{length($Sel$) $=$ length($E_c$)}{
        $G \gets$ AddNewEntity($G, v_n$), $DB\gets$AddNew($DB,v_n$)\;
    }
    \Else{
        \lIf{length($Sel$) $ =1$}{$v_{sel}\gets Sel[0]$}
        \lElse{$v_{sel}\gets$ LLMSelect($Sel$)}
        $G \gets$ MergeEntity($G, v_n, v_{sel}$), $DB\gets$Update($DB,v_{sel}, v_n$)\;
    }
    \Return{$G, DB$}\;
\end{algorithm}

However, conventional ER methods are computationally expensive. 
They are often designed for batch processing across multiple data sources (commonly referred to as dirty ER), aiming to ensure accurate entity resolution by finding all possible matching pairs~\cite{christophides2020overview}.
This process typically requires finding the transitive closure of all detected matches. That is, to definitively merge multiple entities (e.g., A, B, and C) as the same concept, the system must ideally compare all possible pairs (``A-B'', ``A-C'', and ``B-C'') to confirm their equivalence.
This can lead to a quadratic ($O(n^2)$) number of pairwise comparisons, a process that becomes prohibitively slow and computationally expensive when relying on LLMs for high-accuracy judgments.

To address this, we employ a gradient-based ER method, operating on a single document (simplified as the clean ER), which performs ER incrementally as each new entity $v_n$ is extracted. 
This transforms the quadratic batch problem into a simpler, repeated lookup task: determining where the single new entity $v_n$ fits among the already-processed entities in the database.
This incremental process yields two distinct, observable scoring patterns when $v_n$ is reranked against its $top\_k$ most relevant candidates:
\begin{itemize}
    \item \textit{Case A: New Entity.} If $v_n$ is a new conceptual entity, its relevance scores against all existing entities will be uniformly low, showing no significant gradient or discriminative pattern.
    \item \textit{Case B: Existing Entity.} If $v_n$ is an alias of an existing entity, its scores will show a high relevance to the true match (or a small set of equivalent aliases). Due to the reranker's inherent discriminative limitations, this initial high-relevance set might occasionally contain multiple similar entities. This high-relevance set is then typically followed by a \textbf{sharp decline} (a large ``gradient'') before transitioning to a \textbf{gradual slope} of irrelevant entities.
\end{itemize}

Our Gradient-based ER algorithm is designed precisely to detect this sharp decline (characteristic of Case B), allowing us to efficiently isolate the high-relevance set. 
Subsequently, an LLM is utilized for finer-grained distinction when multiple similar entities are identified within this set, differentiating it from the ``no gradient'' scenario (Case A) without quadratic comparisons.

Algorithm \ref{alg:er} shows the above entity resolution process.
For a new entity $v_n$, we first retrieve its $top\_k$ candidates $E_c$ from the vector database $DB$, which are then reranked by $\mathcal{R}$ against $v_n$ and sorted based on their scores $\mathcal{S}$ (Lines 1-3).
We initialize the selection set $Sel$ with the top-scoring candidate $E_c[0]$ and set the initial \texttt{score} to $\mathcal{S}[0]$ (Line 4).
We then iterate through the remaining sorted candidates (Lines 5-8).
The core logic checks if the current score $\mathcal{S}[v_c]$ is still within the gradient threshold $g$ of the previous score (i.e., $\mathcal{S}[v_c] > \texttt{score}/g$).
If the score drop is gentle (passes the check), the candidate $v_c$ is added to $Sel$, and \texttt{score} is updated (Lines 7-8); otherwise, the loop breaks (Line 8) as soon as a sharp score drop is detected.
Finally, the algorithm makes its decision (Lines 9-14).
If the selection set $Sel$ is identical to $E_c$, this indicates that all candidates passed the gradient check.
This corresponds to \textit{Case A}, where the scores lacked discriminative power (i.e., $v_n$ is equally dissimilar to all candidates), so $v_n$ is added as a new entity (Line 9-10).
Conversely, if a gradient was found (i.e., $length(Sel) < length(E_c)$), this signals \textit{Case B}.
We then select the canonical entity $v_{sel}$ from $Sel$, using an LLM (Line 13) if the reranker identifies multiple aliases, and merge $v_n$ with it (Lines 12-14). 
The updated $G$ and $DB$ are then returned (Line 15).

For instance, considering the example in Figure~\ref{fig:overview}, when the new entity $e_9$ is processed, it is first compared with existing entities in the KG.
As depicted in the similarity curve (orange line), $e_9$ shows high similarity with $e_7$, followed by a sharp decline in similarity with other entities like $e_6$, $e_8$, and $e_5$. 
Our gradient-based selection process identifies $e_7$ as the unique, high-confidence match for $e_9$. 
Consequently, $e_9$ is merged with $e_7$, enriching the KG with consolidated information as shown in the final merged entity $e'_7$.

\textbf{Graph-Tree Link (GT-Link).}
The GT-Link $M$ is formalized to complete the BookIndex $B=(T, G, M)$.
As described in the \textit{KG Construction} phase, the origin tree node $n_i$ is recorded for every newly extracted entity $v_i$.
GT-Link is then refined during entity resolution: when an entity $v_n$ is merged into a canonical entity $v_{sel}$, the origin node set of $v_{sel}$ is updated to include all origin nodes previously associated with $v_n$.
This aggregation process creates the final mapping $M: V \to \mathcal{P}(N)$, which bi-directionally links the entities in $G$ to the set of their structural locations (nodes) in $T$.

\section{Agent-based Retrieval}
\label{sec:bookrag}

Real-world document queries are often complex, necessitating operations like modal type filtering, semantic selection, and multi-hop reasoning.
To address this, we propose an agent-based approach in {\modelname}, which intelligently plans and executes operations on the BookIndex.
We first introduce the overall workflow and present two core mechanisms: \textbf{Agent-based Planning}, which formulates the strategy, and the \textbf{Structured Execution}, which includes the retrieval process under the principles of IFT and generation.

\begin{table*}[b]
    \centering
    \setlength{\abovecaptionskip}{0.1cm}
\caption{Three common query categories addressed in {\modelname}.}
    \begin{tabularx}{\textwidth}{l >{\raggedright\arraybackslash\hsize=1.1\hsize}X >{\raggedright\arraybackslash\hsize=1.15\hsize}X >{\raggedright\arraybackslash\hsize=0.75\hsize}X} 
        \toprule
        Query Category  & Description & Core Task  & Example Query  \\
        \midrule
        \textbf{Single-hop} 
        & Queries with a single, distinct information target.
        & \textbf{Scent-based Retrieval}: Retrieve content related to a specific entity or section. 
        & \textit{What is the definition of Information Scent?} \\
        \addlinespace 
        
        \textbf{Multi-hop} 
        & Queries that require synthesizing information from multiple blocks, often by decomposing into sub-problems.
        & \textbf{Decomposing \& Merging}: Decompose into sub-problems, retrieve for each, and synthesize the final answer.
        & \textit{How does Transformer differ from RNNs in handling long-range dependencies?} \\
        \addlinespace 

        \makecell[lt]{\textbf{Global} \\ \textbf{Aggregation}}
        & Queries that require filtering across the entire document and performing calculations.
        & \textbf{Filter \& Aggregation}: Apply filters across the document \& perform aggregation operations (e.g., Count, Sum).
        & \textit{How many figures related to IFT are in Section 4?} \\
        \bottomrule
    \end{tabularx}
    \label{tab:query}
\end{table*}

\subsection{Overall Workflow} 
\label{sec:bookrag-workflow}
\begin{figure}[tb]
    \centering
    \includegraphics[width=\linewidth]{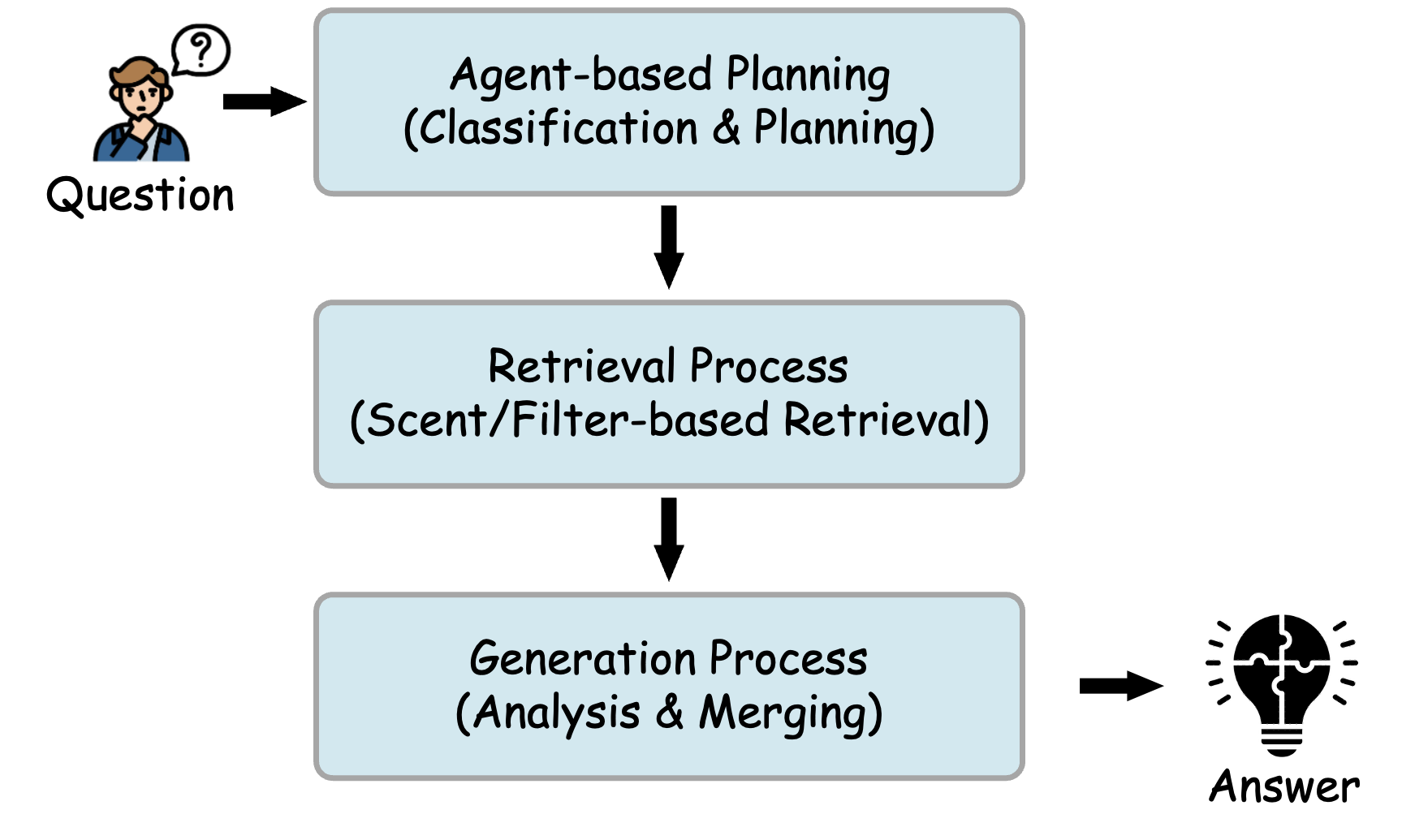}
    \caption{The general workflow of agent-based retrieval in {\modelname}, which contains agent-based planning, retrieval, and generation processes.}
    \label{fig:bookrag-workflow}
\end{figure}

The overall workflow of agent-based retrieval, illustrated in Figure~\ref{fig:bookrag-workflow}, follows a three-stage pipeline designed to address users' queries systematically. 

\textbf{1. Agent-based Planning.}
{\modelname} first performs \textit{Classification \& Plan}. 
This stage aims to distinguish simple keyword-based queries from reasoning questions that require decomposition and analysis. 
For instance, a query like ``How does Transformer differ from RNNs in handling long-range dependencies?'' cannot be solved by retrieving from a single keyword.
Therefore, the planning stage first performs \textbf{query classification}.
Based on this classification and a predefined set of operators designed for the BookIndex, it generates a specific \textbf{operators plan} that effectively guides the retrieval and generation strategies.

\textbf{2. Retrieval Process.}
Guided by the operator plan, the retrieval process executes \textit{Scent/Filter-based Retrieval}. 
This stage navigates the BookIndex $B=(T, G, M)$, either utilizing a \textbf{scent-based retrieval principle} (e.g., following relevant entities in $G$) to find information, or employing various filters (e.g., modal type) to refine the selection.
After reasoning, {\modelname} gets the retrieval set of highly relevant information blocks from the BookIndex.

\textbf{3. Generation Process.}
Finally, all retrieved information enters the generation stage for \textit{Analysis \& Merging}. This stage synthesizes these (often fragmented) pieces of evidence, performs final analysis, and formulates a coherent response.

\subsection{Agent-based Planning}
\label{sec:agent-planning}

The planning stage is the core of {\modelname}, designed to intelligently navigate our BookIndex $B=(T, G, M)$.
To support flexible retrieval, we define four types of operators: Formulator, Selector, Reasoner, and Synthesizer.
These operators can be arbitrarily combined to form tailored execution pipelines, each with adjustable parameters.
{\modelname} dynamically configures and assembles these operators to adapt to the specific requirements of different query categories.
This process involves two sequential steps: first, the agent performs \textit{Query Classification} to determine the appropriate solution strategy, then generates a specific \textit{Operator Plan}.

\begin{figure*}[htp]
    \centering
    \includegraphics[width=\linewidth]{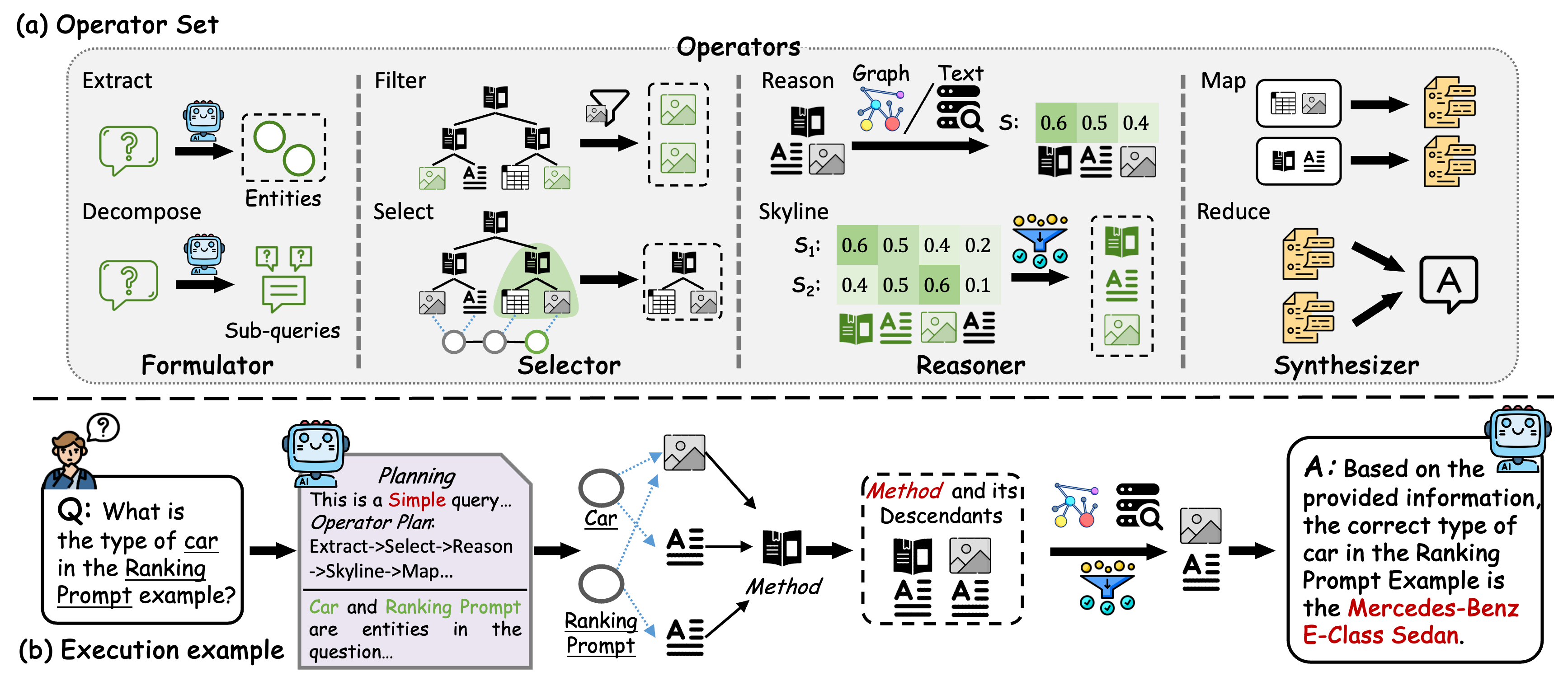}
    \caption{The {\modelname} Operator Library and an Execution Example from MMLongBench dataset:
    (a) a visual depiction of the four operator types (Formulator, Selector, Reasoner, and Synthesizer) and (b) an execution trace for a ``Single-hop'' query, demonstrating the agent-based planning and step-by-step operator execution.}
    \label{fig:bookrag-online}
\end{figure*}

$\bullet$ \textbf{Query Classification}.
To enable agent strategy selection, we focus on three representative query categories defined by their intrinsic complexity and operational demands (Table~\ref{tab:query}): \textit{Single-hop}, \textit{Multi-hop}, and \textit{Global Aggregation}.
This classification is crucial because each category requires a different solution strategy.
For instance, a \textit{Single-hop} query typically requires a single piece of information retrieved via a \textit{Scent-based Retrieval} operation.
In contrast, a \textit{Global Aggregation} query often necessitates analyzing content under multiple filtering conditions, usually involving a sequence of \textit{Filter \& Aggregation} operations across various parts of the document.
Furthermore, {\modelname} is designed to be extensible, allowing for the resolution of a broader range of query types by integrating additional operators.

$\bullet$ \textbf{BookIndex Operators}.
To execute the strategies identified by classification, we designed a set of operators ($\mathcal{O}$) tailored for the BookIndex $B=(T, G, M)$.
These operators, visually depicted in Figure~\ref{fig:bookrag-online}(a) and detailed in Table~\ref{tab:operators}, define the set of operations the agent can employ for diverse query categories.
We group them into four types, which we describe in sequence:


\noindent {\Large \ding{182}} {\textit{Formulator.}}
These are LLM-based operators that prepare the query for execution.
This category includes \texttt{Decompose}, which breaks a \textit{Complex} query into a set of simpler, actionable sub-queries $Q_s$. 
It also includes \texttt{Extract}, which employs an LLM to identify key entities $E_q$ from the query text and link them to corresponding entities in the KG, $G$:
\begin{align}
    Q_s &= \text{LLM}(P_{Dec}, q) = \{q_1, q_2, \dots, q_k\} \label{eq:decompose} \\
    E_q &= \text{LLM}(P_{Ext}, q) = \{e_1, e_2, \dots, e_m\} \label{eq:extract}
\end{align}
Here, $q$ is the original user query, while $P_{Dec}$ and $P_{Ext}$ represent the prompts used to guide the LLM for the decomposition and extraction tasks, respectively.

\noindent {\Large \ding{183}} {\textit{Selector.}} 
%
These operators filter or select specific content ranges from the BookIndex.
\texttt{Filter\_Modal} and \texttt{Filter\_Range} directly apply the explicit constraints $C$ (e.g., modal types, page ranges) generated during the plan.
Operating on the \textit{Tree} $T = (N, E_T)$, these operators produce a filtered subset $N_f$ where the predicate $C(n)$ holds true for each node:
\begin{equation}
    N_f = \{ n \in N \mid C(n) \}
\end{equation}

In contrast, \texttt{Select\_by\_Entity} and \texttt{Select\_by\_Section} target contiguous document segments by retrieving subtrees rooted at specific section nodes.
This process first identifies a set of target section nodes $S_{\text{target}} \subset N$ at a specified depth, where $S_{\text{target}}$ consists of sections either linked to entities $E_q$ via the \textit{GT-Link} $M$ or selected by the LLM.
It then retrieves all descendants of these targets to form the selected node set $N_s$:
\begin{equation}
    N_s = \bigcup_{s \in S_{\text{target}}} \text{Subtree}(s)
\end{equation}

\begin{table*}[b]
    \centering
\caption{Operators utilized in our {\modelname}, categorized by their function.}
\begin{tabularx}{\textwidth}{l >{\raggedright\arraybackslash\hsize=0.35\hsize}X >{\raggedright\arraybackslash\hsize=1.85\hsize}X >{\raggedright\arraybackslash\hsize=0.8\hsize}X}
        \toprule
        \textbf{Operator} & \textbf{Type} & \textbf{Description} & \textbf{Parameters} \\
        \midrule
        \texttt{Decompose} & Formulator & Decompose a complex query into simpler, actionable sub-queries. & \texttt{(Self-contained)} \\
        \texttt{Extract} & Formulator & Identify and extract key entities from the query (links to $G$). & \texttt{(Self-contained)} \\
        \midrule
        
        \texttt{Filter\_Modal} & Selector & Filter retrieved nodes by their modal type (e.g., Table, Figure). & \texttt{modal\_type: str} \\
        \texttt{Filter\_Range} & Selector & Filter nodes based on a specified range (e.g., pages, section). & \texttt{range: (start, end)} \\
        \texttt{Select\_by\_Entity} & Selector & Selects all tree nodes ($N$) in sections linked to a given entity ($V$). & \texttt{entity\_name: str} \\
        \texttt{Select\_by\_Section} & Selector & Uses an LLM to select relevant sections and selects all tree nodes ($N$) within them. & \texttt{query: str, sections: List[str]} \\
        \midrule
        
        \texttt{Graph\_Reasoning} & Reasoner & Performs multi-hop reasoning on subgraph ($G'$) and score tree nodes ($N$) using graph importance and GT-links. & \texttt{start\_entity: str, subgraph: $G'$} \\
        \texttt{Text\_Reasoning} & Reasoner & Rerank retrieved tree nodes ($N$) based on the relevance. & \texttt{query: str} \\
        \texttt{Skyline\_Ranker} & Reasoner & Rerank nodes based on multiple criteria. & \texttt{criteria: List[str]} \\
        \midrule
        \texttt{Map} & Synthesizer &  Uses partially retrieved information to generate a partial answer. & \texttt{(Input: List[str])} \\
        \texttt{Reduce} & Synthesizer & Synthesizes the final answer from partial information or all sub-problem answers. & \texttt{(Input: List[str])} \\
        \bottomrule
\end{tabularx}

    \label{tab:operators}
\end{table*}

\noindent {\Large \ding{184}} {\textit{Reasoner.}} 
%
These operators analyze and refine selected tree nodes. 
\texttt{Graph\_Reasoning} performs multi-hop inference on a subgraph $G'(V',E')$ (extracted from selected nodes $N_s$) starting from entity $e$.
Starting from the retrieved entities, it computes an entity importance vector $I_G \in \mathbb{R}^{|V'|}$ over the subgraph $G'$ using the PageRank algorithm~\cite{haveliwala2002topic}.
These entity scores are then mapped to the tree nodes via the GT-Link matrix $M$ to derive the final tree node importance scores vector $S_G \in \mathbb{R}^{|N_s|}$:
\begin{align}
    I_G &= \text{PageRank}(G', e) \\
    S_G &= I_G \times M
\end{align}
\texttt{Text\_Ranker} evaluates the semantic relevance of the tree node's content to the query $q$, assigning a relevance score $S_T$ to each node.
\texttt{Skyline\_Ranker} employs the Skyline operator to filter nodes based on these multiple criteria (e.g., $S_G$ and $S_T$), retaining only those nodes that are not dominated by any others in terms of the specified scoring dimensions.

\noindent {\Large \ding{185}} {\textit{Synthesizer.}} 
These operators are responsible for content generation. \texttt{Map} performs analysis on specific retrieved information segments to generate partial responses. \texttt{Reduce} synthesizes a final coherent answer by aggregating information from multiple sources, such as partial answers or a collection of retrieved evidence.

$\bullet$ \textbf{Operator Plan}.
After classifying the query ($q$) into its category ($c$), the agent's final task is to generate an executable plan $P$. 
This plan is a specific sequence of operators $\langle o_1, \dots, o_n \rangle$ selected from our library $\mathcal{O}$ with parameters dynamically instantiated based on $q$.
This process is formulated as:
\begin{equation}
    P = \text{Agent}_{\text{Plan}}(q, c, \mathcal{O})
\end{equation}
The plan follows a structured workflow tailored to each category: 
\begin{itemize}
    \item \textit{Single-hop}: The agent first attempts to \texttt{Extract} an entity. If successful, it executes a ``scent-based'' selection; otherwise, it falls back to a section-based strategy. Both paths then proceed to standard reasoning and generation, denoted as $P_{\text{std}}$.
    \begin{align}
        P_{\text{s}}& = 
        \begin{cases}
            \texttt{Extract} \xrightarrow{\text{success}} \texttt{Select\_by\_Entity} \to P_{\text{std}} \\
            \texttt{Extract} \xrightarrow{\text{fail}} \texttt{Select\_by\_Section} \to P_{\text{std}}
        \end{cases} \\
        P_{\text{std}}& = (\texttt{Graph} \parallel \texttt{Text}) \to \texttt{Skyline} \to \texttt{Reduce}
    \end{align}

    \item \textit{Complex}: The agent first decomposes the problem, applies the Single-hop workflow $P_{\text{s}}$ to each sub-problem, and finally synthesizes the results.
    \begin{equation}
        P_{\text{complex}} = \texttt{Decompose} \to P_{\text{s}} \to \texttt{Map} \to \texttt{Reduce}
    \end{equation}

    \item \textit{Global Aggregation}: The workflow involves applying a sequence of filters followed by synthesis.
    \begin{equation}\small
        P_{\text{global}} = \prod (\texttt{Filter\_Modal} \mid \texttt{Filter\_Range}) \to \texttt{Map} \to \texttt{Reduce}
    \end{equation}
    Here, the symbol $\prod$ denotes the nested composition of filters, applying either a modal or range filter at each step.
\end{itemize}

\subsection{Structured Execution}
\label{sec:execution}

Following the planning stage, {\modelname} executes the generated workflow $P$.
This execution phase embodies the cognitive principles of Information Foraging Theory (IFT), effectively translating abstract textual queries into concrete operations.
Specifically, the \texttt{Selector} operators mirror the act of ``navigating to information patches,'' narrowing the vast document space down to relevant scopes.
Subsequently, the \texttt{Reasoner} operators perform ``sense-making within patches,'' where they analyze and refine the information within these focused scopes.
Finally, the \texttt{Synthesizer} generates the answer based on the processed evidence.
This design minimizes the cost of attention by ensuring computational resources are focused solely on high-value data patches.

\paragraph{Scent/Filter-based Retrieval}
The execution begins by narrowing the scope.
Aligning with IFT, \texttt{Selector} operators identify relevant ``patches'' by following ``information scents'' (e.g., key entities in question) or applying explicit filter constraints.
This process reduces the full node set $N$ to a focused node subset $N_s$:
\begin{equation}
    N_s = \texttt{Selector}(N, \text{params}_{\text{sel}})
\end{equation}

This pre-selection minimizes noise and ensures that subsequent reasoning is applied only to highly relevant contexts, optimizing the foraging cost.
Subsequently, within this focused scope, \texttt{Reasoner} operators evaluate nodes using multiple dimensions, such as graph topology and semantic relevance.
We then employ the \texttt{Skyline\_Ranker} to get the final retrieval set.
Unlike fixed top-$k$ retrieval, the Skyline operator retains the \textit{Pareto frontier} of nodes, retaining nodes that are valuable in at least one dimension while discarding dominated ones:
\begin{equation}
    N_R = \texttt{Skyline\_Ranker}(\{S_G(n), S_T(n) \mid n \in N_s\})
\end{equation}

\paragraph{Analysis \& Merging Generation}
In the final stage, the \texttt{Synthesizer} operator generates the coherent answer by aggregating the refined evidence:
\begin{equation}
    A = \texttt{Synthesizer}(q, N_R)
\end{equation}
The \texttt{Map} operator performs fine-grained analysis on individual evidence blocks or sub-problems (from \texttt{Decompose}) to generate intermediate insights.
The \texttt{Reduce} operator then aggregates these partial results, such as answers to decomposed sub-queries or statistical counts from a global filter, to construct the final response.
This separation ensures that the system can handle both detailed content extraction and high-level reasoning synthesis effectively.

To illustrate this end-to-end process, Figure~\ref{fig:bookrag-online}(b) presents an execution trace for a ``Single-hop'' query: ``What is the type of \textit{car} in the \textit{Ranking Prompt} example?''.
In the planning phase, the agent classifies the query and generates a specific workflow.
Subsequently, it identifies key entities (e.g., ``car'') via \texttt{Extract}, retrieves relevant nodes via \texttt{Select\_by\_Entity}, refines them through reasoning and Skyline filtering, and finally synthesizes the answer using \texttt{Reduce}.


\section{Experiments}
\label{sec:exp}

In our experiments, we evaluate {\modelname} against several strong baseline methods, with an in-depth comparison of their efficiency and accuracy on document QA tasks.

\subsection{Setup}

\begin{table}[h]
    \centering
\caption{Datasets used in our experiments. EM and F1 denote Exact Match and F1-score, respectively.}
    \begin{tabular}{c|ccc}
        \toprule
        Dataset & MMLongBench & M3DocVQA & Qasper  \\
        \midrule
        Questions & 669 & 633 & 640  \\
        Documents & 85 & 500 & 192 \\
        Avg. Pages & 42.16 & 8.52 & 10.95 \\
        Avg. Images & 25.92 & 3.51 & 3.43 \\
        Tokens & 2,816,155  & 3,553,774 & 2,265,349 \\
        Metrics & EM, F1 & EM, F1 & Accuracy, F1 \\
        \bottomrule
    \end{tabular}
    \label{tab:dataset}
\end{table}

\paragraph{Datasets \& Question Synthesis.}
We use three widely adopted benchmarking datasets for complex document QA tasks: 
MMLongBench~\cite{ma2024mmlongbench}, M3DocVQA~\cite{cho2024m3docrag}, and Qasper~\cite{dasigi2021dataset}.
MMLongBench is a comprehensive benchmark designed to evaluate QA capabilities on long-form documents, covering diverse categories such as guidebooks, financial reports, and industry files.
M3DocVQA is an open-domain benchmark designed to test RAG systems on a diverse collection of HTML-type documents sourced from Wikipedia pages\footnote{https://www.wikipedia.org/}.
Qasper is a QA dataset focused on scientific papers, where questions require retrieving evidence from the entire document. 
We filtered the datasets to remove documents with low clarity or incoherent structures.
To address the scarcity of global-level questions in the original benchmarks, we synthesize additional QA pairs by having an LLM generate global questions from selected document elements (e.g., tables or figures).
These questions are then answered and meticulously refined by human annotators via an outsourcing process, with this additional QA pairs constituting less than 20\% of our final QA pairs. 
The statistics of these datasets are presented in Table \ref{tab:dataset}.

\begin{table*}[htb]
    \centering    
    \caption{Performance comparison of different methods across various datasets for solving complex document QA tasks. The best and second-best results are marked in bold and underlined, respectively.}
    \begin{tabular}{ll cccccc}
        \toprule
        \multirow{2}{*}{Baseline Type} & \multirow{2}{*}{Method} & \multicolumn{2}{c}{MMLongBench} & \multicolumn{2}{c}{M3DocVQA} & \multicolumn{2}{c}{Qasper} \\
        \cmidrule(lr){3-4} \cmidrule(lr){5-6} \cmidrule(lr){7-8}
& &(Exact Match) & (F1-score) & (Exact Match) & (F1-score) & (Accuracy) & (F1-score) \\
\midrule
\multirow{3}{*}{Conventional RAG} 
& {BM25}            & 18.3 & 20.2 & 34.6 & 37.8 & 38.1 & 42.5   \\
& {Vanilla RAG}     & 16.5 & 18.0 & 36.5 & 40.2 & 40.6 & 44.4  \\
& {Layout + Vanilla}& 18.1 & 19.8 & 36.9 & 40.2 & 40.7 & 44.6  \\
\midrule
\multirow{3}{*}{Graph-based RAG} 
& {RAPTOR}          & 21.3 & 21.8 & 34.3 & 37.3 & 39.4 & 44.1 \\
& {GraphRAG-Local}  & 7.7  & 8.5  & 23.7 & 25.6 & 35.9 & 39.2 \\
& {GraphRAG-Global} & 5.3  & 5.6  & 20.2 & 22.0 & 24.0 & 24.1 \\
\midrule
\multirow{4}{*}{Layout segmented RAG}
& {MM-Vanilla}      & 6.8  & 8.4  & 25.1 & 27.7 & 27.9 & 29.3 \\
& {Tree-Traverse}   & 12.7 & 14.4 & 33.3 & 36.2 & 27.3 & 32.1 \\
& {GraphRanker}     & 21.2 & 22.7 & \underline{43.0} & \underline{47.8} & 32.9 & 37.6 \\
& {DocETL}          & \underline{27.5} & \underline{28.6} & 40.9 & 43.3 & \underline{42.3} & \underline{50.4} \\
\midrule
\textbf{Our proposed} & \textbf{\modelname} & {\bf 43.8} & {\bf 44.9}& {\bf 61.0} & {\bf 66.2} & {\bf 55.2} & {\bf 61.1}\\
        \bottomrule
    \end{tabular}
    \label{tab:main-res}
\end{table*}

\paragraph{Metrics.}
We adhere to the official metrics specified by each dataset for QA. 
Our primary evaluation relies on Exact Match (EM), accuracy, and token-based F1-score. To assess efficiency, we also measure time cost and token usage during the response phase.
Additionally, for methods including PDF parsing, we also evaluate retrieval recall.
To establish the ground truth for this, we manually label the specific PDF blocks (e.g., texts, titles, tables, images, and formulas) required to answer each question.
This labeling process is guided by the metadata of ground-truth evidence provided in each dataset; we filter candidate blocks using the given modality (all datasets), page numbers (MMLongBench), and evidence statements (Qasper). 
Any blocks that remained non-unique after this filtering process are manually annotated.
In cases where a PDF parsing error made the ground-truth item unavailable, the retrieval recall for that query is recorded as 0.

\paragraph{Baselines.}
Our experiments consider three model configurations:
\begin{itemize}
\item \textbf{Conventional RAG:}
These methods are the most common pipeline for document analysis, where the raw text is first extracted and then chunked into segments of a specified size.
We select strong and widely used retrieval models: BM25~\cite{robertson1994some} and Vanilla RAG. 
We also implement Layout+Vanilla, a variant that uses document layout analysis for semantic chunking.

\item \textbf{Graph-based RAG:}
These methods first extract textual content from documents and then leverage graph data during retrieval. We select {RAPTOR} \cite{sarthi2024raptor} and {GraphRAG}~\cite{edge2024local}.
Specifically, {GraphRAG} has two versions: {GraphRAG-Global} and {GraphRAG-Local}, which employ global and local search methods, respectively.

\item \textbf{Layout segmented RAG:}
This category encompasses methods that utilize layout analysis to segment document content into discrete structural units.
We include: MM-Vanilla, which utilizes multi-modal embeddings for visual and textual content; a tree-based method inspired by PageIndex~\cite{PageIndex}, denoted as TreeTraverse, where an LLM navigates the document's tree structure; 
DocETL~\cite{shankar2024docetl}, a declarative system for complex document processing;
and {GraphRanker}, a graph-based method extended from HippoRAG~\cite{gutierrez2024hipporag} that applies Personalized PageRank~\cite{haveliwala2002topic} to rank the relevant nodes.
\end{itemize}

\paragraph{Implementation details.}
For a fair comparison, both {\modelname} and all baseline methods are powered by a unified set of state-of-the-art (SOTA) and widely adopted backbone models from the Qwen family~\cite{yang2025qwen3,bai2025qwen2,zhang2025qwen3,zhang2024gme}.
We employ MinerU~\cite{wang2024mineru} for robust document layout parsing.
We set the threshold of gradient $g$ as $0.6$, and more details are provided in the appendix of our technical report~\cite{fullVersion}.
Our source code, prompts, and detailed configurations are available at \href{https://github.com/sam234990/\modelname}{github.com/sam234990/\modelname}.




\definecolor{c1}{RGB}{216,174,174} 
\definecolor{c2}{RGB}{224,175,107}  
\definecolor{c3}{RGB}{222,117,123} 
\definecolor{c4}{RGB}{250, 221, 104} 

\definecolor{c5}{RGB}{174,223,172} 
\definecolor{c6}{RGB}{138,170,214}  
\definecolor{c7}{RGB}{248,199,1} 
\definecolor{c8}{RGB}{255,0,127} 

\definecolor{c11}{RGB}{172, 196, 226} 
\definecolor{c12}{RGB}{208, 221, 238} 

\definecolor{c21}{RGB}{163,137,214} 


\definecolor{c9}{RGB}{150,195,125} %
\definecolor{c10}{RGB}{230,189,69} %

\definecolor{c13}{RGB}{115,107,157} %
\definecolor{c14}{RGB}{208,108,157} %

\pgfplotstableread[row sep=\\,col sep=&]{
datasets & BM25 & Vanilla & PDFVanilla & RAPTOR & LGraphRAG & GGraphRAG & MMVanilla & Tree & Graph & DocETL & GBCRAG \\
2 & 2221.35 & 2660.04 & 737.32 & 2465.62 & 2856.48 & 10251.78 & 1914.1 & 4819.7 & 7742.4 & 35795.76 & 23414.56 \\
4 & 1994426 & 2460249 & 722086 & 2155739 & 2119131 & 6518810 & 1356436 & 3124960 & 2434433 & 53219015 & 4992827 \\
6 & 2606.98 & 3051.72 & 1506.16 & 3137.68 & 3068.38 & 11523.92 & 2276.28 & 4938.22 & 7036.71 & 13663.73 & 5461.56 \\
8 & 2365400 & 3175202 & 1544413 & 2835857 & 2638381 & 3569715 & 1498474 & 2789601 & 2924305 & 16214344 & 2644755 \\
10 & 2485.12 & 2029.31 & 1904.44 & 2080.69 & 3849.39 & 10046.34 & 2499.92 & 4889.06 & 7443.2 & 21990.92 & 9601.48 \\
12 & 2124381 & 1921327 & 1540979 & 2475737 & 2463287 & 3210290 & 1672692 & 2483223 & 2242880 & 25482506 & 4021480 \\
}\Qtimetoken

\pgfplotstableread[row sep=\\,col sep=&]{
datasets & zero & cot & BM25 & vanilla & hippo & lightl & lighth & lighthy & graphl & graphg & {\modelname} \\ 
2 & 426105 & 795941 & 5594336 & 5556774 & 6121633 & 14617559 & 15467710 & 18515561 & 1536156 & 605937663 & 12575315 \\ 
4 & 89649 & 172217 & 675569 & 203654 & 886793 & 3088800 & 3223300 & 4817210 & 601000 & 1394436159 & 5191529 \\ 
}\Qtoken

\subsection{Overall results}
\label{sec:overallExp}
In this section, we present a comprehensive evaluation of {\modelname}, analyzing its complex QA performance, retrieval effectiveness, and query efficiency compared to state-of-the-art baselines.

$\bullet$ \textbf{QA Performance of {\modelname}}.
We compare the QA performance of {\modelname} against three categories of baselines, as shown in Table \ref{tab:main-res}.
The results indicate that {\modelname} achieves state-of-the-art performance across all datasets, substantially outperforming the top-performing baseline by 18.0\% in Exact Match on M3DocVQA.
%
Layout + Vanilla consistently outperforms Vanilla RAG, confirming that layout parsing preserves essential structural information for better retrieval.
Besides, the suboptimal results of Tree-Traverse and GraphRanker highlight the limitations of relying solely on hierarchical navigation or graph-based reasoning, which often miss cross-sectional context or drift into irrelevant scopes.
In contrast, {\modelname}'s superiority stems from the synergy of its unified Tree-Graph BookIndex and Agent-based Planning.
By effectively classifying queries and configuring optimal workflows, our {\modelname} overcomes limitations of context fragmentation and static query workflow within existing baselines, ensuring precise evidence retrieval and accurate generation.

\begin{table}[h]
    \centering
\caption{Retrieval recall comparison among layout-based methods. The best and second-best results are marked in bold and underlined, respectively.}
    \begin{tabular}{lccc}
\toprule
Method &  MMLongBench & M3DocVQA & Qasper  \\
\midrule
Layout + Vanilla & 26.3 & 33.8 & \underline{33.5} \\
MM-Vanilla & 7.5 & 19.7 & 14.9 \\
Tree-Traverse & 11.2 & 19.5 & 14.5 \\
GraphRanker & \underline{26.4} & \underline{44.5} & 28.6 \\
\textbf{\modelname} & \textbf{57.6} & \textbf{71.2} & \textbf{63.5}\\
\bottomrule
    \end{tabular}
    \label{tab:recall}
\end{table}

$\bullet$ \textbf{Retrieval performance of {\modelname}.}
To validate our retrieval design, we evaluate the retrieval recall of {\modelname} against other {layout-based baselines} on the ground-truth layout blocks.
The experimental results demonstrate that {\modelname} achieves the highest recall across all datasets, notably reaching 71.2\% on M3DocVQA and significantly outperforming the next best baseline (GraphRanker, max 44.5\%).
This performance advantage stems from our IFT-inspired \textbf{Selector $\to$ Reasoner} workflow: the Agent-based Planning first classifies the query, enabling the Selector to narrow the search to a precise \textit{information patch}, followed by the Reasoner's analysis.
Crucially, after the \texttt{Skyline\_Ranker} process, the average number of retained nodes is 9.87, 6.86, and 8.6 across the three datasets, which is comparable to the standard top-$k$ ($k=10$) setting, ensuring high-quality retrieval without inflating the candidate size.

%
%

\begin{figure}[h]
    \centering
    \quad \ref{eff_leg}\\
    \subfigure[MMLongBench]{
    \begin{tikzpicture}[scale=0.45]
            \begin{axis}[
                ybar=0.5pt,
                bar width=0.6cm,
                width=0.5\textwidth,
                height=0.26\textwidth,
                xtick=data,
                xticklabels={\huge Query Time, \huge Token cost,\huge Query Time, \huge Token cost,\huge Query Time, \huge Token cost},
                legend style={
                 at={(0.0,1.05)}, 
    anchor=south west, 
                    align=left, 
                legend columns=3,
                draw=none},
                legend image code/.code={
                    \draw [#1, line width=0.5pt] (0cm,-0.1cm) rectangle (0.3cm,0.2cm); },
                legend to name=eff_leg,
                xmin=1,xmax=3,
                ymin=100,ymax=100000,
                ytick = {100, 1000, 10000, 100000},
                ymode = log,
                tick align=inside,
                ticklabel style={font=\Huge},
                every axis plot/.append style={line width = 2.5pt},
                every axis/.append style={line width = 2.5pt},
                ylabel={\textbf{\Huge time (s)}}
                ]
\addplot[fill=c1] table[x=datasets,y=BM25]{\Qtimetoken};
\addplot[fill=c2] table[x=datasets,y=Vanilla]{\Qtimetoken};
\addplot[fill=c3] table[x=datasets,y=PDFVanilla]{\Qtimetoken};
\addplot[fill=c4] table[x=datasets,y=RAPTOR]{\Qtimetoken};
\addplot[fill=c5] table[x=datasets,y=LGraphRAG]{\Qtimetoken};
\addplot[fill=c6] table[x=datasets,y=GGraphRAG]{\Qtimetoken};
\addplot[fill=c11] table[x=datasets,y=MMVanilla]{\Qtimetoken};
\addplot[fill=c12] table[x=datasets,y=Tree]{\Qtimetoken};
\addplot[fill=c7] table[x=datasets,y=Graph]{\Qtimetoken};
\addplot[fill=c21] table[x=datasets,y=DocETL]{\Qtimetoken};
\addplot[fill=c8] table[x=datasets,y=GBCRAG]{\Qtimetoken};

\legend{\small {BM25},\small {Vanilla RAG},\small {Layout + Vanilla}, \small {RAPTOR}, \small {GraphRAG-Local}, \small {GraphRAG-Global}, \small {MM-Vanilla}, \small {Tree-Traverse}, \small {GraphRanker}, \small {DocETL}, \small {\modelname}}
            \end{axis}
        \end{tikzpicture}
    \begin{tikzpicture}[scale=0.45]
        \begin{axis}[
            ybar=0.5pt,
            bar width=0.6cm,
            width=0.5\textwidth,
            height=0.26\textwidth,
            xtick=data,
            xticklabels={\huge Query Time, \huge Token cost,\huge Query Time, \huge Token cost,\huge Query Time, \huge Token cost},
            xmin=3,xmax=5,
            ymin=100000,ymax=100000000,
            ytick = {100000, 1000000, 10000000, 100000000},
            ymode = log,
            tick align=inside,
            ticklabel style={font=\Huge},
            every axis plot/.append style={line width = 2.5pt},
            every axis/.append style={line width = 2.5pt},
            ylabel={\textbf{\Huge token (M)}}
            ]
\addplot[fill=c1] table[x=datasets,y=BM25]{\Qtimetoken};
\addplot[fill=c2] table[x=datasets,y=Vanilla]{\Qtimetoken};
\addplot[fill=c3] table[x=datasets,y=PDFVanilla]{\Qtimetoken};
\addplot[fill=c4] table[x=datasets,y=RAPTOR]{\Qtimetoken};
\addplot[fill=c5] table[x=datasets,y=LGraphRAG]{\Qtimetoken};
\addplot[fill=c6] table[x=datasets,y=GGraphRAG]{\Qtimetoken};
\addplot[fill=c11] table[x=datasets,y=MMVanilla]{\Qtimetoken};
\addplot[fill=c12] table[x=datasets,y=Tree]{\Qtimetoken};
\addplot[fill=c7] table[x=datasets,y=Graph]{\Qtimetoken};
\addplot[fill=c21] table[x=datasets,y=DocETL]{\Qtimetoken};
\addplot[fill=c8] table[x=datasets,y=GBCRAG]{\Qtimetoken};
        \end{axis}
    \end{tikzpicture}
    }
    \subfigure[M3DocVQA]{
		\begin{tikzpicture}[scale=0.45]
            \begin{axis}[
                ybar=0.5pt,
                bar width=0.6cm,
                width=0.5\textwidth,
                height=0.26\textwidth,
                xtick=data,
                xticklabels={\huge Query Time, \huge Token cost,\huge Query Time, \huge Token cost,\huge Query Time, \huge Token cost},
                xmin=5,xmax=7,
                ymin=1000,ymax=100000,
                ytick = {1000, 10000, 100000},
                ymode = log,
                tick align=inside,
                ticklabel style={font=\Huge},
                every axis plot/.append style={line width = 2.5pt},
                every axis/.append style={line width = 2.5pt},
                ylabel={\textbf{\Huge time (s)}}
                ]
\addplot[fill=c1] table[x=datasets,y=BM25]{\Qtimetoken};
\addplot[fill=c2] table[x=datasets,y=Vanilla]{\Qtimetoken};
\addplot[fill=c3] table[x=datasets,y=PDFVanilla]{\Qtimetoken};
\addplot[fill=c4] table[x=datasets,y=RAPTOR]{\Qtimetoken};
\addplot[fill=c5] table[x=datasets,y=LGraphRAG]{\Qtimetoken};
\addplot[fill=c6] table[x=datasets,y=GGraphRAG]{\Qtimetoken};
\addplot[fill=c11] table[x=datasets,y=MMVanilla]{\Qtimetoken};
\addplot[fill=c12] table[x=datasets,y=Tree]{\Qtimetoken};
\addplot[fill=c7] table[x=datasets,y=Graph]{\Qtimetoken};
\addplot[fill=c21] table[x=datasets,y=DocETL]{\Qtimetoken};
\addplot[fill=c8] table[x=datasets,y=GBCRAG]{\Qtimetoken};
            \end{axis}
        \end{tikzpicture}
        \begin{tikzpicture}[scale=0.45]
            \begin{axis}[
                ybar=0.5pt,
                bar width=0.6cm,
                width=0.5\textwidth,
                height=0.26\textwidth,
                xtick=data,
                xticklabels={\huge Query Time, \huge Token cost,\huge Query Time, \huge Token cost,\huge Query Time, \huge Token cost},
                xmin=7,xmax=9,
                ymin=100000,ymax=50000000,
                ytick = {100000, 1000000, 10000000},
                ymode = log,
                tick align=inside,
                ticklabel style={font=\Huge},
                every axis plot/.append style={line width = 2.5pt},
                every axis/.append style={line width = 2.5pt},
                ylabel={\textbf{\Huge token (M)}}
                ]
\addplot[fill=c1] table[x=datasets,y=BM25]{\Qtimetoken};
\addplot[fill=c2] table[x=datasets,y=Vanilla]{\Qtimetoken};
\addplot[fill=c3] table[x=datasets,y=PDFVanilla]{\Qtimetoken};
\addplot[fill=c4] table[x=datasets,y=RAPTOR]{\Qtimetoken};
\addplot[fill=c5] table[x=datasets,y=LGraphRAG]{\Qtimetoken};
\addplot[fill=c6] table[x=datasets,y=GGraphRAG]{\Qtimetoken};
\addplot[fill=c11] table[x=datasets,y=MMVanilla]{\Qtimetoken};
\addplot[fill=c12] table[x=datasets,y=Tree]{\Qtimetoken};
\addplot[fill=c7] table[x=datasets,y=Graph]{\Qtimetoken};
\addplot[fill=c21] table[x=datasets,y=DocETL]{\Qtimetoken};
\addplot[fill=c8] table[x=datasets,y=GBCRAG]{\Qtimetoken};
            \end{axis}
        \end{tikzpicture}
	}
    \subfigure[Qasper]{
		\begin{tikzpicture}[scale=0.45]
            \begin{axis}[
                ybar=0.5pt,
                bar width=0.6cm,
                width=0.5\textwidth,
                height=0.26\textwidth,
                xtick=data,
                xticklabels={\huge Query Time, \huge Token cost,\huge Query Time, \huge Token cost,\huge Query Time, \huge Token cost},
                xmin=9,xmax=11,
                ymin=1000,ymax=50000,
                ytick = {1000, 10000},
                ymode = log,
                tick align=inside,
                ticklabel style={font=\Huge},
                every axis plot/.append style={line width = 2.5pt},
                every axis/.append style={line width = 2.5pt},
                ylabel={\textbf{\Huge time (s)}}
                ]
\addplot[fill=c1] table[x=datasets,y=BM25]{\Qtimetoken};
\addplot[fill=c2] table[x=datasets,y=Vanilla]{\Qtimetoken};
\addplot[fill=c3] table[x=datasets,y=PDFVanilla]{\Qtimetoken};
\addplot[fill=c4] table[x=datasets,y=RAPTOR]{\Qtimetoken};
\addplot[fill=c5] table[x=datasets,y=LGraphRAG]{\Qtimetoken};
\addplot[fill=c6] table[x=datasets,y=GGraphRAG]{\Qtimetoken};
\addplot[fill=c11] table[x=datasets,y=MMVanilla]{\Qtimetoken};
\addplot[fill=c12] table[x=datasets,y=Tree]{\Qtimetoken};
\addplot[fill=c7] table[x=datasets,y=Graph]{\Qtimetoken};
\addplot[fill=c21] table[x=datasets,y=DocETL]{\Qtimetoken};
\addplot[fill=c8] table[x=datasets,y=GBCRAG]{\Qtimetoken};
            \end{axis}
        \end{tikzpicture}
        \begin{tikzpicture}[scale=0.45]
            \begin{axis}[
                ybar=0.5pt,
                bar width=0.6cm,
                width=0.5\textwidth,
                height=0.26\textwidth,
                xtick=data,
                xticklabels={\huge Query Time, \huge Token cost,\huge Query Time, \huge Token cost,\huge Query Time, \huge Token cost},
                xmin=11,xmax=13,
                ymin=1000000,ymax=50000000,
                ytick = {1000000, 10000000},
                ymode = log,
                tick align=inside,
                ticklabel style={font=\Huge},
                every axis plot/.append style={line width = 2.5pt},
                every axis/.append style={line width = 2.5pt},
                ylabel={\textbf{\Huge token (M)}}
                ]
\addplot[fill=c1] table[x=datasets,y=BM25]{\Qtimetoken};
\addplot[fill=c2] table[x=datasets,y=Vanilla]{\Qtimetoken};
\addplot[fill=c3] table[x=datasets,y=PDFVanilla]{\Qtimetoken};
\addplot[fill=c4] table[x=datasets,y=RAPTOR]{\Qtimetoken};
\addplot[fill=c5] table[x=datasets,y=LGraphRAG]{\Qtimetoken};
\addplot[fill=c6] table[x=datasets,y=GGraphRAG]{\Qtimetoken};
\addplot[fill=c11] table[x=datasets,y=MMVanilla]{\Qtimetoken};
\addplot[fill=c12] table[x=datasets,y=Tree]{\Qtimetoken};
\addplot[fill=c7] table[x=datasets,y=Graph]{\Qtimetoken};
\addplot[fill=c21] table[x=datasets,y=DocETL]{\Qtimetoken};
\addplot[fill=c8] table[x=datasets,y=GBCRAG]{\Qtimetoken};
            \end{axis}
        \end{tikzpicture}
	}
    \caption{Comparison of query efficiency.}
    \label{fig:query-efficiency}
\end{figure}
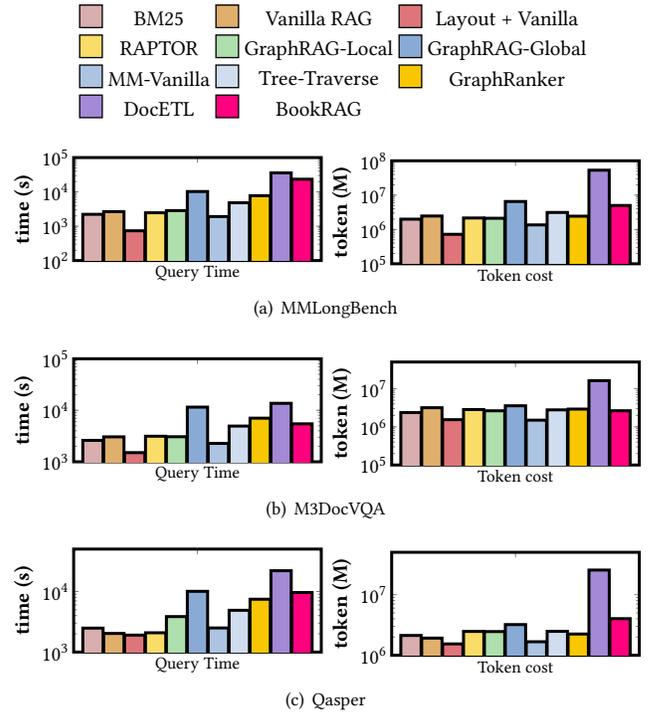

$\bullet$ \textbf{Efficiency of {\modelname}.}
We further evaluate the efficiency in terms of query time and token consumption, as illustrated in Figure \ref{fig:query-efficiency}.
Overall, {\modelname} maintains time and token costs comparable to existing Graph-based RAG methods.
While purely text-based RAG approaches generally exhibit lower latency and token usage due to the absence of VLM processing for images, {\modelname} maintains a balanced efficiency among multi-modal methods.
In terms of token usage, {\modelname} reduces consumption by an order of magnitude compared to the strongest baseline, DocETL.
Notably, on the MMLongBench dataset, DocETL consumes over 53 million tokens, whereas {\modelname} requires less than 5 million.
Regarding the query latency, our method also achieves a speedup of up to $2\times$ compared to DocETL.

\pgfplotstableread[row sep=\\,col sep=&]{
topk & Accuracy & Recall \\
1 & 63.2 & 37 \\
3 & 66.1 & 36.5 \\
5 & 68.8 & 37.2 \\
7 & 63.4 & 37.1 \\
9 & 67.4 & 37 \\
}\topkmultihop
\pgfplotstableread[row sep=\\,col sep=&]{
topk & Accuracy & Recall \\
1 & 61.5 & 65.6 \\
3 & 61.2 & 65 \\
5 & 65.4 & 69.2 \\
7 & 63.6 & 67.3 \\
9 & 47.7 & 52.5 \\
}\topkhotpot

\pgfplotstableread[row sep=\\,col sep=&]{
datasets & EM & F1 \\
2 & 47.6 & 47.6 \\
4 & 34.1 & 37 \\
6 & 54.6 & 54.9 \\
}\mmlongtype
\pgfplotstableread[row sep=\\,col sep=&]{
datasets & EM & F1 \\
2 & 52.9 & 64.2 \\
4 & 51.8 & 55.3 \\
6 & 68.4 & 65.4 \\
}\qaspertype

\subsection{Detailed Analysis}

In this section, we provide a more in-depth examination of our {\modelname}. 
We first conduct an ablation study to validate the contribution of each component, followed by an experiment on the impact of gradient-based ER and QA performance across different query types.
Furthermore, we perform a comprehensive error analysis, compare the effectiveness of our entity resolution method, and present a case study.

$\bullet$ \textbf{Ablation study.}
To evaluate the contribution of each core component in {\modelname}, we design several variants by removing specific components:
\begin{itemize}
    \item w/o Gradient ER: Replaces the gradient-based entity resolution with a Basic ER by merging the same-name entities.
    \item w/o Planning: Removes the Agent-based Planning, defaulting to a static, standard workflow for all queries.
    \item w/o \texttt{Selector}: Removes the Selector operators, forcing Reasoners to score all candidate nodes.
    \item w/o \texttt{Graph\_Reasoning}: Removes the \texttt{Graph\_Reasoning} operator. Consequently, the \texttt{Skyline\_Ranker} is also disabled as scoring becomes single-dimensional.
    \item w/o \texttt{Text\_Reasoning}: Removes the \texttt{Text\_Reasoning} operator. Similarly, the \texttt{Skyline\_Ranker} is disabled, relying solely on graph-based scores.
\end{itemize}

\begin{table}[htb]
    \centering
\caption{Comparing the QA performance of different variants of {\modelname}. EM and F1 denote Exact Match and F1-score, respectively.}
    \begin{tabular}{lcccc}
        \toprule
\multirow{2}{*}{Method variants} &  \multicolumn{2}{c}{MMLongBench} & \multicolumn{2}{c}{Qasper} \\
\cmidrule(lr){2-3} \cmidrule(lr){4-5}
        & EM & F1 & Accuracy & F1\\
\midrule
{\modelname} (Full) & 43.8 & 44.9 & 55.2 & 61.1 \\
\quad w/o gradient ER & 40.1 & 42.8 & 48.9 & 57.3 \\
\quad w/o Planning & 30.8 & 33.2 & 40.9 & 48.5 \\
\quad w/o \texttt{Selector} & 42.5 & 43.1 & 52.5 & 59.1 \\
\quad w/o \texttt{Graph\_Reasoning} & 39.8 & 41.5 & 51.4 & 58.4 \\
\quad w/o \texttt{Text\_Reasoning} & 39.0 & 40.3 & 47.2 & 52.5 \\
\bottomrule
    \end{tabular}
    \label{tab:variant}
\end{table}

The first variant evaluates the impact of KG quality on retrieval performance. The second and third variants assess the necessity of our Agent-based Planning and IFT-inspired selection mechanism, respectively.
Finally, the last two variants validate the effectiveness of our multi-dimensional reasoning and dynamic Skyline filtering strategy.
As shown in Table \ref{tab:variant}, the performance degradation across all variants confirms the essential role of each module in {\modelname}.
Specifically, the performance drop in the \textit{w/o Gradient ER} variant highlights the critical role of a high-quality, connectivity-rich KG in supporting effective reasoning.
Removing the \textit{Planning} mechanism results in the most significant performance loss, confirming that a static workflow is insufficient for handling diverse types of queries.
The \textit{w/o Selector} variant, while maintaining competitive accuracy, incurs a prohibitive computational cost ($>2\times$ tokens on Qasper), validating the efficiency of our IFT-inspired "narrow-then-reason" strategy.
%

\pgfplotstableread[row sep=\\,col sep=&]{
metric & Basic & ER & label \\
2    & 1.0 & 0.875 & 1327 \\
4    & 1.0 & 1.257 & 3.6E-3 \\
6    & 1.0 & 0.899 & 14.8 \\
8    & 1.0 & 0.787 & 169 \\
}\mmlongbenchdata

\pgfplotstableread[row sep=\\,col sep=&]{
metric & Basic & ER & label \\
2    & 1.0 & 0.893 & 531 \\
4    & 1.0 & 1.227 & 5.4e-3 \\
6    & 1.0 & 0.910 & 15.0 \\
8    & 1.0 & 0.826 & 106 \\
}\qasperdata

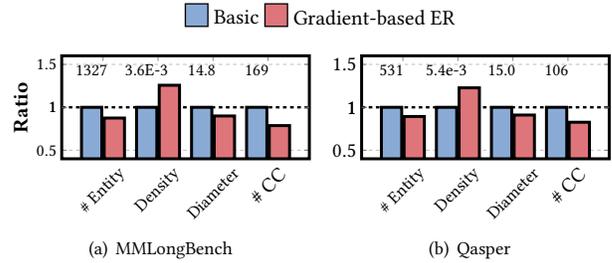
\begin{figure}[h]
    \centering
    
    \quad \ref{named_legend} \\
    
    \subfigure[MMLongBench]{
        \begin{tikzpicture}[scale=0.45]
            \begin{axis}[
                ybar,
                bar width=0.6cm,       
                width=0.5\textwidth,
                height=0.26\textwidth,
                xtick=data,	
                xticklabels={\huge \# Entity, \huge Density, \huge Diameter, \# CC},
                xticklabel style={rotate=45, font=\Huge, align=center}, 
                ymin=0.4, ymax=1.6,
                xmin=0.5, xmax=9.5,
                ytick={0.5, 1.0, 1.5},
                yticklabel style={font=\huge}, 
                ylabel={\textbf{\Huge Ratio}}, 
                ymajorgrids=true,
                grid style=dashed,
                tick align=inside,
                legend to name=named_legend,
                legend style={
                anchor=north,legend columns=4,
                draw=none},
                legend image code/.code={
                    \draw [#1, line width=0.5pt] (0cm,-0.1cm) rectangle (0.3cm,0.2cm); 
                },
                extra y ticks = {1.0},
                extra y tick style={grid=major, grid style={line width=2pt, black, dashed}}, 
                nodes near coords style={
                    font=\huge,      
                    color=black,
                    anchor=south,    
                    yshift=0.8cm     
                },
                every axis plot/.append style={line width = 2.5pt},
                every axis/.append style={line width = 2.5pt}
            ]
            
            \addplot[
                fill=c6, 
                draw=black, 
                nodes near coords=\pgfplotspointmeta, 
                point meta=explicit symbolic
            ] table[x=metric, y=Basic, meta=label]{\mmlongbenchdata};
            
            \addplot[fill=c3, draw=black] table[x=metric, y=ER]{\mmlongbenchdata};
\legend{\small {{Basic}},\small {Gradient-based ER}}
            
            \end{axis}
        \end{tikzpicture}
    }
    \subfigure[Qasper]{
        \begin{tikzpicture}[scale=0.45]
            \begin{axis}[
                ybar,
                bar width=0.6cm,
                width=0.5\textwidth,
                height=0.26\textwidth,
                xtick=data,	
                xticklabels={\huge \# Entity, \huge Density, \huge Diameter, \# CC},
                xticklabel style={rotate=45, font=\Huge, align=center}, 
                ymin=0.4, ymax=1.6,
                xmin=0.5, xmax=9.5,
                ytick={0.5, 1.0, 1.5},
                yticklabel style={font=\Huge},
                tick align=inside,
                ymajorgrids=true,
                grid style=dashed,
                extra y ticks = {1.0},
                extra y tick style={grid=major, grid style={line width=2pt, black, dashed}}, 
                nodes near coords style={
                    font=\huge,
                    color=black,
                    anchor=south,
                    yshift=0.8cm
                },
                every axis plot/.append style={line width = 2.5pt},
                every axis/.append style={line width = 2.5pt}
            ]
            
            \addplot[
                fill=c6, 
                draw=black,
                nodes near coords=\pgfplotspointmeta,
                point meta=explicit symbolic
            ] table[x=metric, y=Basic, meta=label]{\qasperdata};
            
            \addplot[fill=c3, draw=black] table[x=metric, y=ER]{\qasperdata};
            
            \end{axis}
        \end{tikzpicture}
    }
    
    \caption{Comparison of graph statistics. Values are normalized to the Basic setting (Baseline=1.0). Absolute values for Basic are annotated.
    Note that density values are abbreviated (e.g., 3.6E-3 denotes $3.6 \times 10^{-3}$).
    }
    \label{fig:graph_stats}
\end{figure}

$\bullet$ \textbf{Impact of Gradient-based Entity Resolution.}
To evaluate the quality of our constructed KG, we compare the graph statistics of our Gradient-based ER against a Basic KG construction.
The Basic setting employs simple exact name matching for entity merging, which is standard practice in many graph-based methods.
Figure~\ref{fig:graph_stats} presents the comparative results, normalizing the metrics (Entity count, Density, Diameter of the Largest Connected Component, and Number of Connected Components) against the Basic baseline.
The results demonstrate that our Gradient-based ER significantly optimizes KG.
Specifically, it reduces the number of entities (by 12\%) while substantially boosting graph density (by over 20\% across datasets).
This structural shift indicates that our ER module effectively identifies the same conceptual entities that possess different names.
Consequently, the resulting graphs are more compact and cohesive, as evidenced by the reduced diameter and fewer connected components, which mitigates graph fragmentation and facilitates better connectivity for graph reasoning.

\begin{figure}[h]
    \centering
    \quad \ref{quality_ch}\\
    \subfigure[MMLongBench]{
    \begin{tikzpicture}[scale=0.45]
            \begin{axis}[
                ybar=0.5pt,
                bar width=0.6cm,
                width=0.5\textwidth,
                height=0.26\textwidth,
                xtick=data,	
                xticklabels={\huge Single, \huge Multi, \huge Global},
                ytick={20, 50, 80},
                yticklabel style={font=\Huge},
                legend style={
                anchor=north,legend columns=4,
                draw=none},
                legend image code/.code={
                    \draw [#1, line width=0.5pt] (0cm,-0.1cm) rectangle (0.3cm,0.2cm); },
                legend to name=quality_ch,
                xmin=1,xmax=7,
                ymin=0,ymax=100,
                tick align=inside,
                ticklabel style={font=\Huge},
                every axis plot/.append style={line width = 2.5pt},
                every axis/.append style={line width = 2.5pt},
                ylabel={\textbf{{\Huge Score}}}
                ]
\addplot[fill=c6] table[x=datasets,y=EM]{\mmlongtype};
\addplot[fill=c3] table[x=datasets,y=F1]{\mmlongtype};
\legend{\small {EM / Accuracy},\small {F1-score}}
            \end{axis}
        \end{tikzpicture}
    }
    \subfigure[Qasper]{
		\begin{tikzpicture}[scale=0.45]
            \begin{axis}[
                ybar=0.5pt,
                bar width=0.6cm,
                width=0.5\textwidth,
                height=0.26\textwidth,
                xtick=data,	
                xticklabels={\huge Single, \huge Multi, \huge Global},
                ytick={20, 50, 80},
                yticklabel style={font=\Huge},
                legend style={
                anchor=north,legend columns=4,
                draw=none},
                legend image code/.code={
                    \draw [#1, line width=0.5pt] (0cm,-0.1cm) rectangle (0.3cm,0.2cm); },
                xmin=1,xmax=7,
                ymin=0,ymax=100,
                tick align=inside,
                ticklabel style={font=\Huge},
                every axis plot/.append style={line width = 2.5pt},
                every axis/.append style={line width = 2.5pt},
                ylabel={\textbf{{\Huge Score}}}
                ]
\addplot[fill=c6] table[x=datasets,y=EM]{\qaspertype};
\addplot[fill=c3] table[x=datasets,y=F1]{\qaspertype};
            \end{axis}
        \end{tikzpicture}
	}
    \caption{QA performance breakdown by different query types (Single-hop, Multi-hop, and Global). The blue bars represent Exact Match (EM) for MMLongBench and Accuracy for Qasper, while the red bars represent the F1-score.}
    \label{fig:qa_performance}
\end{figure}
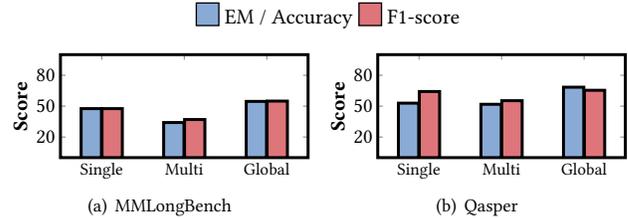

$\bullet$ \textbf{QA performance under different query types.}
Figure~\ref{fig:qa_performance} breaks down the performance of {\modelname} across Single-hop, Multi-hop, and Global aggregation query types.
We observe that Multi-hop queries generally present a greater challenge compared to Single-hop ones, resulting in a slight performance decrease.
This trend reflects the inherent difficulty of retrieving and reasoning over disjoint pieces of evidence.
It further validates our agent-based planning strategy, which handles different query types separately.

%

\begin{figure*}[htbp]
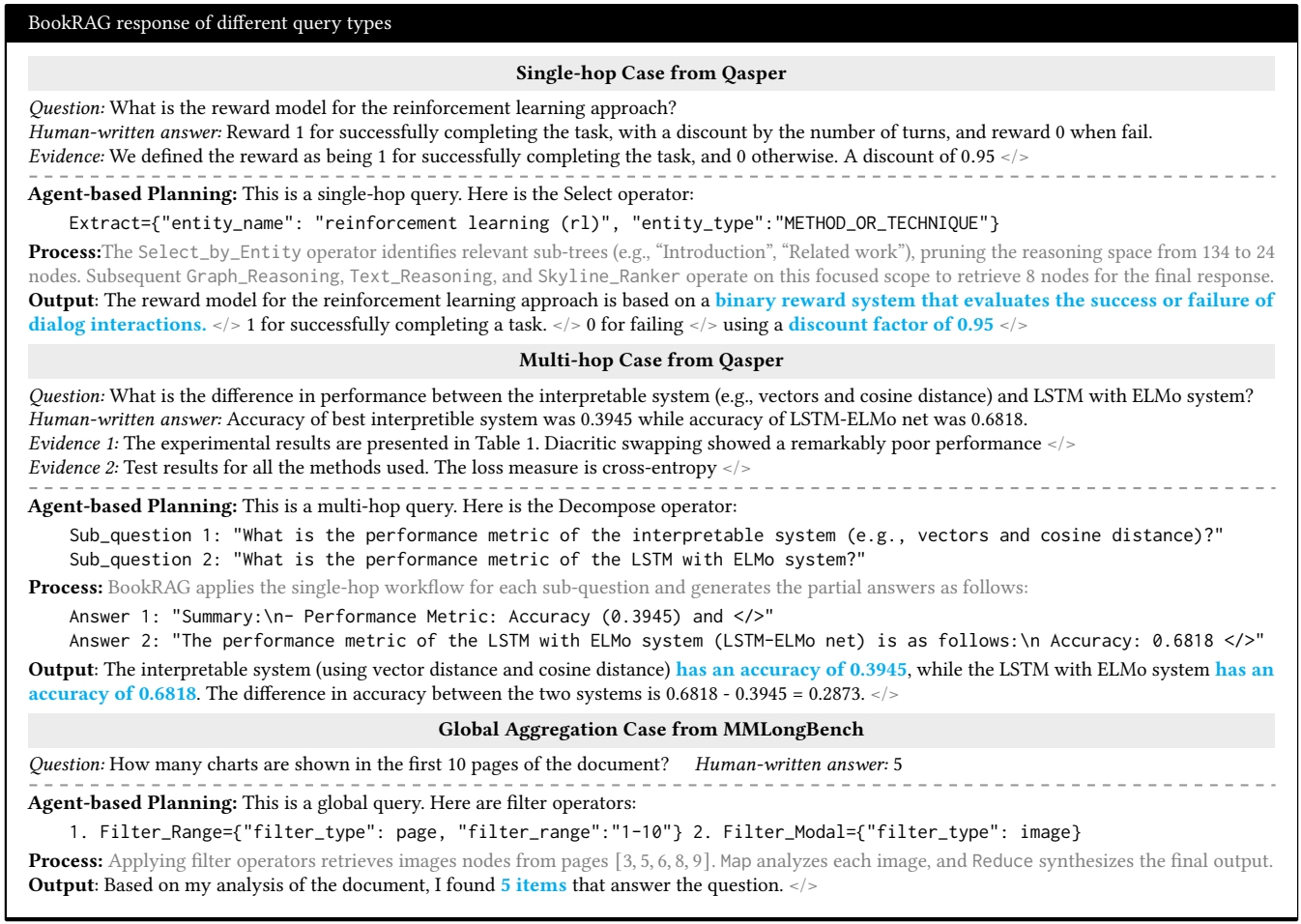

    \centering
    \small
    \linespread{0.94}\selectfont
    \begin{tcolorbox}[ragstyle, title={\modelname} response of different query types]
\vspace{-0.2cm}
\caseheader{Single-hop Case from Qasper}

\textit{Question:} What is the reward model for the reinforcement learning approach?\\
\textit{Human-written answer:} Reward 1 for successfully completing the task, with a discount by the number of turns, and reward 0 when fail.\\
\textit{Evidence:} We defined the reward as being 1 for successfully completing the task, and 0 otherwise. A discount of $0.95$ \textcolor{gray}{</>}
\vspace{-0.2cm}
\intrasep 
\vspace{-0.1cm}
\textbf{Agent-based Planning:} This is a single-hop query. Here is the Select operator:
\begin{verbatim}
    Extract={"entity_name": "reinforcement learning (rl)", "entity_type":"METHOD_OR_TECHNIQUE"}
\end{verbatim}
\textbf{Process:}\textcolor{gray}{ The \texttt{Select\_by\_Entity} operator identifies relevant sub-trees (e.g., ``Introduction'', ``Related work''), pruning the reasoning space from 134 to 24 nodes. Subsequent \texttt{Graph\_Reasoning}, \texttt{Text\_Reasoning}, and \texttt{Skyline\_Ranker} operate on this focused scope to retrieve 8 nodes for the final response.}
\\
\textbf{Output}: The reward model for the reinforcement learning approach is based on a \textcolor{cyan}{\bf binary reward system that evaluates the success or failure of dialog interactions.} \textcolor{gray}{</>} 1 for successfully completing a task. \textcolor{gray}{</>} 0 for failing \textcolor{gray}{</>} using a \textcolor{cyan}{\bf discount factor of 0.95} \textcolor{gray}{</>}

\caseheader{Multi-hop Case from Qasper}
\textit{Question:} What is the difference in performance between the interpretable system (e.g., vectors and cosine distance) and LSTM with ELMo system?\\
\textit{Human-written answer:} Accuracy of best interpretible system was 0.3945 while accuracy of LSTM-ELMo net was 0.6818.\\
\textit{Evidence 1:} The experimental results are presented in Table 1. Diacritic swapping showed a remarkably poor performance \textcolor{gray}{</>}\\
\textit{Evidence 2:} Test results for all the methods used. The loss measure is cross-entropy  \textcolor{gray}{</>}
\vspace{-0.2cm}
\intrasep 
\vspace{-0.1cm}
\textbf{Agent-based Planning:} This is a multi-hop query. Here is the Decompose operator:
\begin{verbatim}
    Sub_question 1: "What is the performance metric of the interpretable system (e.g., vectors and cosine distance)?"
    Sub_question 2: "What is the performance metric of the LSTM with ELMo system?"
\end{verbatim}
\textbf{Process:} \textcolor{gray}{{\modelname} applies the single-hop workflow for each sub-question and generates the partial answers as follows:}
\begin{verbatim}
    Answer 1: "Summary:\n- Performance Metric: Accuracy (0.3945) and </>"
    Answer 2: "The performance metric of the LSTM with ELMo system (LSTM-ELMo net) is as follows:\n Accuracy: 0.6818 </>"
\end{verbatim}

\textbf{Output}: The interpretable system (using vector distance and cosine distance) \textcolor{cyan}{\bf has an accuracy of 0.3945}, while the LSTM with ELMo system \textcolor{cyan}{\bf has an accuracy of 0.6818}. The difference in accuracy between the two systems is 0.6818 - 0.3945 = 0.2873.  \textcolor{gray}{</>}

    \caseheader{Global Aggregation Case from MMLongBench}
    
\textit{Question:} How many charts are shown in the first 10 pages of the document? 
\quad
\textit{Human-written answer:} 5

\vspace{-0.2cm}
\intrasep 
\vspace{-0.1cm}

\textbf{Agent-based Planning:} This is a global query. Here are filter operators:
\begin{verbatim}
    1. Filter_Range={"filter_type": page, "filter_range":"1-10"} 2. Filter_Modal={"filter_type": image}
\end{verbatim}
\textbf{Process:} \textcolor{gray}{Applying filter operators retrieves images nodes from pages $[3,5,6,8,9]$. \texttt{Map} analyzes each image, and \texttt{Reduce} synthesizes the final output.}\\
\textbf{Output}: Based on my analysis of the document, I found \textcolor{cyan}{\bf 5 items} that answer the question. \textcolor{gray}{</>}
\end{tcolorbox}
\vspace{-2em}
\caption{Case study of responses across different query types from MMLongBench and Qasper. {\bf \textcolor{cyan}{CYAN TEXT}} highlights correct content generated by {\modelname}. \textcolor{gray}{GRAY TEXT} describes the internal process, and \textcolor{gray}{</>} marks omitted irrelevant parts.}
    \label{fig:case}
\end{figure*}

\begin{figure}[tbp]
    \centering
    \subfigure[MMLongBench]{
        \includegraphics[width=0.46\linewidth]{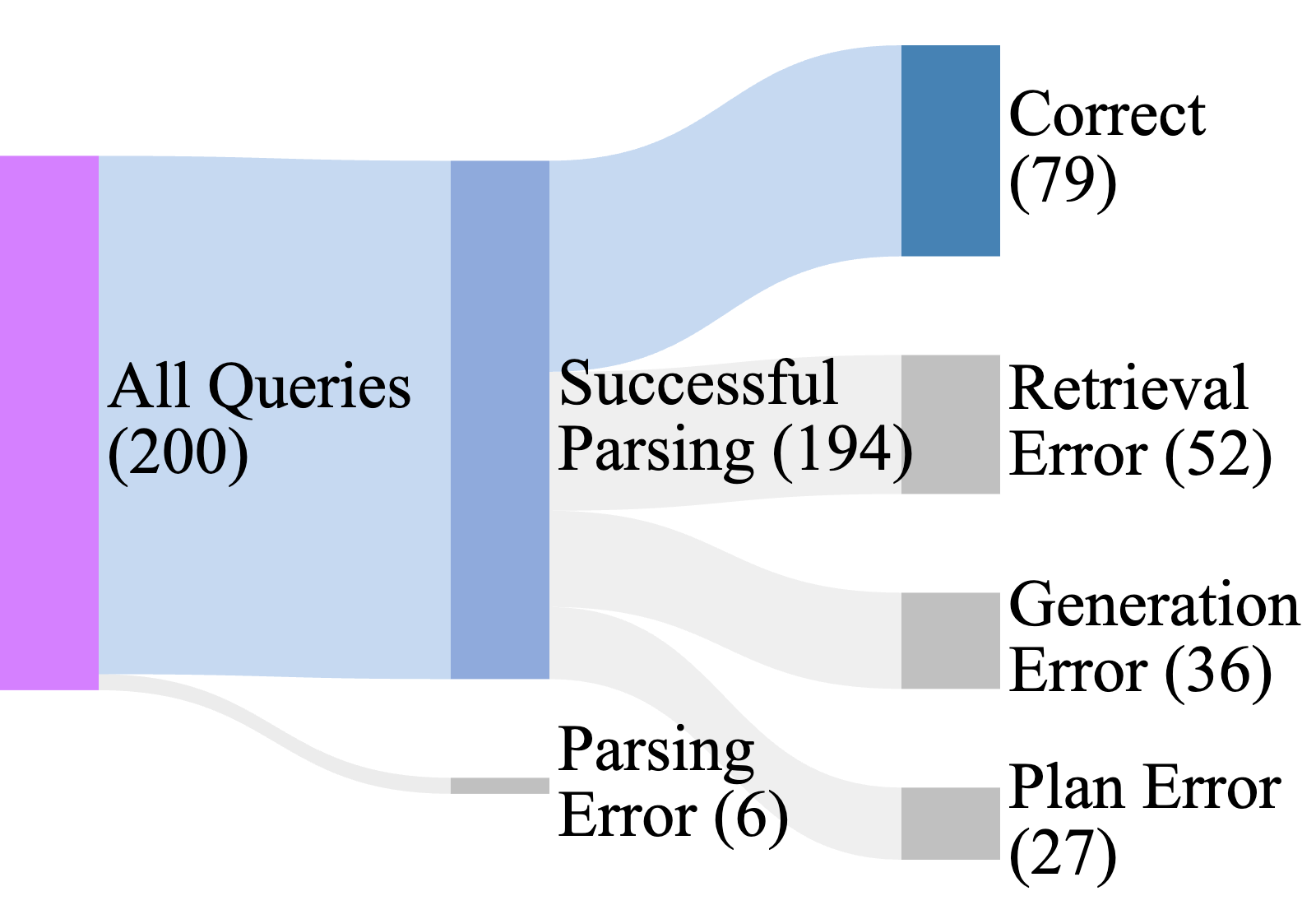}
    }
	\subfigure[Qasper]{
        \includegraphics[width=0.46\linewidth]{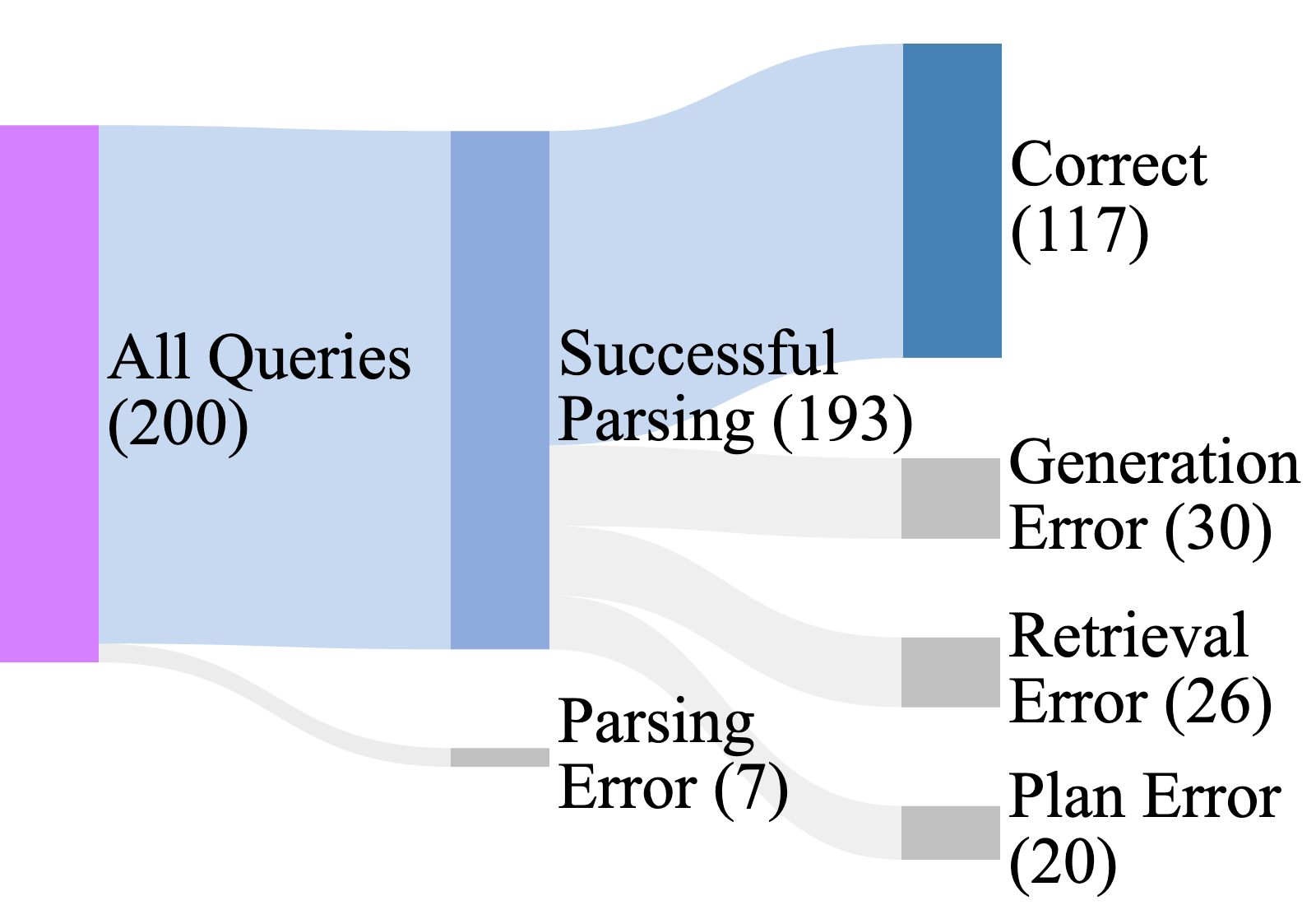}
	}
    \caption{Error analysis on 200 sampled queries from MMLongBench and Qasper datasets.}
    \label{fig:err_analysis}
\end{figure}

$\bullet$ \textbf{Error Response analysis.}
To diagnose the performance bottlenecks of {\modelname}, we conduct a fine-grained error analysis on 200 sampled queries from each dataset, tracing the error propagation as shown in Figure~\ref{fig:err_analysis}.
We categorize failures into four types: PDF Parsing, Plan, Retrieval, and Generation errors.
The results identify Retrieval Error as the dominant failure mode, followed by Generation Error, reflecting the persistent challenge of locating and synthesizing multimodal evidence.
Regarding Plan Error, our qualitative analysis reveals a specific failure pattern: the planner tends to over-decompose detailed single-hop queries into unnecessary multi-hop sub-tasks.
This fragmentation leads to disjointed retrieval paths, effectively preventing the model from synthesizing a cohesive final answer from the scattered sub-responses.

$\bullet$ \textbf{Case study.} 
Figure~\ref{fig:case} illustrates {\modelname}'s answering workflow across Single-hop, Multi-hop, and Global queries.
The results demonstrate that by leveraging specific operators (\texttt{Select}, \texttt{Decompose}, and \texttt{Filter}), {\modelname} effectively prunes search spaces.
For example, in the Single-hop case, the reasoning space is significantly reduced from 134 to 24 nodes. 
This capability allows the system to efficiently isolate relevant evidence from noise, ensuring precise answer generation.

\section{Conclusion}
\label{sec:conclusion}

In this paper, we propose {\modelname}, a novel method built upon Book Index, a document-native, structured Tree-Graph index specifically designed to capture the intricate relations of structural documents.
By employing an agent-based method to dynamically configure retrieval and reasoning operators, our approach achieves state-of-the-art performance on multiple benchmarks, demonstrating significant superiority over existing baselines in both retrieval precision and answer accuracy.
In the future, we will explore an integrated document-native database system that supports data formatting, knowledge extraction, and intelligent querying.

\clearpage
\balance
\bibliographystyle{ACM-Reference-Format}
\bibliography{sample-base,references}

\appendix
\clearpage

\section{Experimental details}

\subsection{Evaluation Metrics}
\label{appendix:metrics}

In this section, we provide the detailed definitions and calculation procedures for the metrics used in our main experiments.

\subsubsection{Answer Extraction and Normalization}
Standard RAG models typically generate free-form natural language responses, which may contain extraneous conversational text (e.g., ``The answer is...''). Directly comparing these raw outputs with concise ground truth labels (e.g., ``Option A'' or ``12.5'') can lead to false negatives. 

Following official evaluation protocols, we employ an LLM-based extraction step to align the model output with the ground truth format before calculation. Let $y_{raw}$ denote the raw response generated by the RAG system and $y_{gold}$ denote the ground truth. We define the extracted answer $\hat{y}$ as:
\begin{equation}
    \hat{y} = \text{LLM}_{\text{extract}}(y_{raw}, \text{Instruction})
\end{equation}
where $\text{LLM}_{\text{extract}}$ extracts the key information (e.g., the key entity for span extraction) from $y_{raw}$. We further apply standard normalization $\mathcal{N}(\cdot)$ (e.g., lowercasing, removing punctuation) to both $\hat{y}$ and $y_{gold}$.
\subsubsection{QA Performance Metrics}
Based on the ground truth $y_{gold}$ and the model's response (either raw $y_{raw}$ or extracted $\hat{y}$), we compute the following metrics:

\paragraph{Accuracy (Inclusion-based).}
Following prior works~\cite{schick2024toolformer,mallen2022not,asai2023self}, we utilize accuracy as a soft-match metric. We consider a prediction correct if the normalized gold answer is included in the model's generated response, rather than requiring a strict exact match. This accounts for the uncontrollable nature of LLM generation.
\begin{equation}
    \text{Accuracy} = \frac{1}{N} \sum_{i=1}^{N} \mathbb{I}(\mathcal{N}(y_{gold, i}) \subseteq \mathcal{N}(y_{raw, i}))
\end{equation}
where $\subseteq$ denotes the substring inclusion relation.

\paragraph{Exact Match (EM).}
Unlike accuracy, Exact Match is a strict metric. It measures whether the normalized \textit{extracted} answer $\hat{y}$ is character-for-character identical to the ground truth.
\begin{equation}
    \text{EM} = \frac{1}{N} \sum_{i=1}^{N} \mathbb{I}(\mathcal{N}(\hat{y}_i) = \mathcal{N}(y_{gold, i}))
\end{equation}

\paragraph{F1-score.}
For questions requiring text span answers, we utilize the token-level F1-score between the extracted answer $\hat{y}$ and the ground truth $y_{gold}$. Treating them as bags of tokens $T_{\hat{y}}$ and $T_{gold}$:
\begin{equation}
    P = \frac{|T_{\hat{y}} \cap T_{gold}|}{|T_{\hat{y}}|}, \quad R = \frac{|T_{\hat{y}} \cap T_{gold}|}{|T_{gold}|}, \quad \text{F1} = \frac{2 \cdot P \cdot R}{P + R}
\end{equation}

\subsubsection{Retrieval Recall}
As described in the main text, we evaluate retrieval quality based on the granularity of parsed PDF blocks (e.g., paragraphs, tables, images). 
For a given query $q$, let $\mathcal{B}_{gold}$ be the set of manually labeled ground-truth blocks required to answer $q$, and $\mathcal{B}_{ret}$ be the set of unique blocks retrieved by the system. The Retrieval Recall is defined as:
\begin{equation}
    \text{Recall}_{ret} = \begin{cases} 
    0 & \text{if parsing error occurs on } \mathcal{B}_{gold} \\
    \frac{|\mathcal{B}_{ret} \cap \mathcal{B}_{gold}|}{|\mathcal{B}_{gold}|} & \text{otherwise}
    \end{cases}
\end{equation}
Specifically, if a ground-truth block is lost due to PDF parsing failures (i.e., it does not exist in the candidate pool), it is considered strictly unretrievable, resulting in a recall contribution of 0 for that specific block.

\subsection{Implementation details}
\label{sec:imp}

We implement {\modelname} in Python, utilizing MinerU~\cite{wang2024mineru} for robust document layout parsing.
For a fair comparison, both {\modelname} and all baseline methods are powered by a unified set of state-of-the-art (SOTA) and widely adopted backbone models from the Qwen family~\cite{yang2025qwen3,bai2025qwen2,zhang2025qwen3,zhang2024gme}, including LLM, vision-language model (VLM), and embedding models.
Specifically, we utilize Qwen3-8B~\cite{yang2025qwen3} as the default LLM, Qwen2.5VL-30B~\cite{bai2025qwen2} as the vision-language model (VLM), Qwen3-Embedding-0.6B~\cite{zhang2025qwen3} for text embedding, gme-Qwen2-VL-2B-Instruct~\cite{zhang2024gme} for multi-modal embedding, and Qwen3-Reranker-4B~\cite{zhang2025qwen3} for reranking.
We primarily select models under the 10B parameter scale to balance efficiency and effectiveness.
However, for the VLM, we adopt the 30B version, as the 8B counterpart exhibited significant performance deficits, frequently failing to answer correctly even when provided with ground-truth images.
All experiments were conducted on a Linux operating system running on a high-performance server equipped with an Intel Xeon 2.0GHz CPU, 1024GB of memory, and 8 NVIDIA GeForce RTX A5000 GPUs, each with 24 GB of VRAM.
Specifically, to ensure a fair comparison of efficiency, all methods were executed serially, and the reported time costs reflect this sequential processing mode.
For methods involving document chunking and retrieval ranking, we standardize the chunk size at 500 tokens and set the retrieval top-$k$ to 10 to ensure consistent candidate pool sizes across baselines.
For further reproducibility, our source code and detailed implementation configurations are publicly available at our repository: \href{https://github.com/sam234990/BookRAG}{https://github.com/sam234990/BookRAG}.

\subsection{Prompts}

Specifically, we present the prompts designed for agent-based query classification (Figure~\ref{fig:prompt_query}), question decomposition (Figure~\ref{fig:prompt_decom}), and filter operator generation (Figure~\ref{fig:prompt_filter}). Additionally, we illustrate the prompt employed for entity resolution judgment (Figure~\ref{fig:prompt_er}) during the graph construction phase.

\begin{figure*}[t]
\begin{PromptBox}
You are an expert query analyzer. Your only task is to classify the user's question into one of three categories: "simple", "complex", or "global". Respond only with the specified JSON object.

Category Definitions:
1.  single-hop: The question can be fully answered by retrieving information from a SINGLE, contiguous location in the document (e.g., one specific paragraph, one complete table, or one figure).
    - This includes questions that require reasoning or comparison, as long as all the necessary data is present within that single retrieved location.
    - Example: "What is the title of Figure 2?"
    - Example: "How do 5% of the Latinos see economic upward mobility for their children?" -> This is SIMPLE because the answer can be found by looking at a single chart or paragraph.

2.  multi-hop: The question requires decomposition into multiple simple sub-questions, where each sub-question must be answered by a separate retrieval action.
    - It often contains a nested or indirect constraint that requires a preliminary step to resolve before the main question can be answered.
    - Example: "What is the color of the personality vector...?" -> This is COMPLEX because it requires two separate retrieval actions.

3.  global: The question requires an aggregation operation (e.g., counting, listing, summarizing) over a set of items that are identified by a clear structural filter.
    - Example: "How many tables are in the document?" -> This is GLOBAL because the process is to filter for all items of type 'table'.

User Query: (*@\textbf{query}@*)

\end{PromptBox}
\caption{The prompt for query classification.}
\label{fig:prompt_query}
\end{figure*}

\begin{figure*}[t]
\begin{PromptBox}
You are a query decomposition expert. You have been given a "complex" question. Your task is to break it down into a series of simple, atomic sub-questions and classify each one by type.

**Crucial Instructions:**
1.  Each `retrieval` sub-question MUST be a direct information retrieval task that can be answered independently by looking up a specific fact, number, or value in the document.
2.  **`retrieval` sub-questions MUST NOT depend on the answer of another sub-question.** They should be parallelizable. All logic for combining their results must be placed in a final `synthesis` question.
3.  A `synthesis` question requires comparing, calculating, or combining the answers of the previous `retrieval` questions. It does **NOT** require a new lookup in the document.

You MUST provide your response in a JSON object with a single key 'sub_questions', which contains a list of objects. Each object must have a 'question' (string) and a 'type' (string: "retrieval" or "synthesis").

--- EXAMPLE 1 (Correct Decomposition with Independent Lookups) ---
Complex Query: "What is the color of the personality vector in the soft-labled personality embedding matrix that with the highest Receptiviti score for User A2GBIFL43U1LKJ?"

Expected JSON Output:
{{
  "sub_questions": [
    {{"question": "What are all the Receptiviti scores for each personality vector for User A2GBIFL43U1LKJ?",
      "type": "retrieval"}},
    {{"question": "What is the mapping of personality vectors to their colors in the soft-labled personality embedding matrix?",
      "type": "retrieval"}},
    {{"question": "From the gathered scores, identify the personality vector with the highest score, and then find its corresponding color from the vector-to-color mapping.",
      "type": "synthesis"}}
  ]
}}
--- END EXAMPLE 1 ---

--- EXAMPLE 2 (Decomposition with retrieval and synthesis steps) ---
Complex Query: "According to the report, which one is greater in population in the survey? Foreign born Latinos, or the Latinos interviewed by cellphone?"

Expected JSON Output:
{{
  "sub_questions": [
    {{"question": "According to the report, what is the population of foreign born Latinos in the survey?",
      "type": "retrieval"}},
    {{"question": "According to the report, what is the population of Latinos interviewed by cellphone in the survey?",
      "type": "retrieval"}},
    {{"question": "Which of the two population counts is greater?",
      "type": "synthesis"}}
  ]
}}
--- END EXAMPLE 2 ---

Now, perform the decomposition for the following query.

User Query: (*@\textbf{query}@*)

\end{PromptBox}
\caption{The prompt for query decomposition.}
\label{fig:prompt_decom}
\end{figure*}

\begin{figure*}[t]
\begin{PromptBox}
You are a highly specialized AI assistant. Your only function is to analyze a "Global Query" and return a single, valid JSON object that specifies both the filtering steps and the final aggregation operation. You MUST NOT output any other text or explanation.

### INSTRUCTIONS \& DEFINITIONS ###

1.  **Filters**: You MUST determine the list of `filters` to apply. Even if the filter is for the whole document (e.g., all tables), the `filters` list must be present.
    - `filter_type`: One of ["section", "image", "table", "page"].
        - `section`: Use for structural parts like chapters, sections, appendices, or references.
        - `image`: Use for visual elements like figures, images, pictures, or plots.
        - `table`: Use for tabular data.
        - `page`: Use for specific page numbers or ranges.
    - `filter_value`: (Optional) Can be provided for "section" (e.g., a section title) or "page" (e.g., '3-10' or '5'). **For "image" or "table", this value MUST be null.**

2.  **Operation**: Determine the final aggregation operation.
    - `operation`: One of ["COUNT", "LIST", "SUMMARIZE", "ANALYZE"].

### EXAMPLES OF YOUR TASK ###

User: "How many figures are in this paper from Page 3 to Page 10?"
Assistant: {{"filters": [{{"filter_type": "page", "filter_value": "3-10"}}, {{"filter_type": "image"}}], "operation": "COUNT"}}

User: "Summarize the discussion about 'data augmentation' in the 'Methodology' section."
Assistant: {{"filters": [{{"filter_type": "section", "filter_value": "Methodology"}}], "operation": "SUMMARIZE"}}

User: "How many chapters are in this report?"
Assistant: {{"filters": [{{"filter_type": "section"}}], "operation": "COUNT"}}

### YOUR CURRENT TASK ###

User: "{query}"
User Query: (*@\textbf{query}@*)

\end{PromptBox}
\caption{The prompt for Filter operator generation.}
\label{fig:prompt_filter}
\end{figure*}

\begin{figure*}[t]
\begin{PromptBox}
-Goal-
You are an expert Entity Resolution Adjudicator. Your task is to determine if a "New Entity" refers to the exact same real-world concept as one of the "Candidate Entities" provided from a knowledge graph. Your output must be a JSON object containing the ID of the matching candidate (or -1) and a brief explanation for your decision.
-Context-
You will be given one "New Entity" recently extracted from a text. You will also be given a list of "Candidate Entities" that are semantically similar, retrieved from an existing knowledge base. Each candidate has a unique `id` for you to reference.
---
-Core Task & Rules-
1.  **Analyze the "New Entity"**: Carefully read its name, type, and description to understand what it is.
2.  **Field-by-Field Adjudication**: To determine a match, you must evaluate each field with a specific focus:
    * **`entity_name` (High Importance):** The names must be extremely similar, a direct abbreviation (e.g., "LLM" vs. "Large Language Model"), or a well-known alias. **If the names represent distinct, parallel concepts (like "Event Detection" and "Named Entity Recognition"), they are NOT a match, even if their descriptions are very similar.**
    * **`entity_type` (Medium Importance):** The types do not need to be identical, but they must be closely related and compatible (e.g., `COMPANY` and `ORGANIZATION` could describe the same entity).
    * **`description` (Contextual Importance):** The descriptions may differ as they are often extracted from different parts of a document. Your task is to look past surface-level text similarity and determine if they fundamentally describe the **same underlying object or concept**.

3.  **Be Strict and Conservative**: Your standard for a match must be very high. An incorrect merge can corrupt the knowledge graph. A missed merge is less harmful.
    * Surface-level similarities are not enough. The underlying concepts must be identical.
    * For example, "Apple" (the fruit) and "Apple Inc." (the company) are NOT a match.
    * **When in doubt, you MUST output -1.**
    * **Assume No Match by Default**: In a large knowledge graph, most new entities are genuinely new. You should start with the assumption that the "New Entity" is unique. You must find **strong, convincing evidence** across all fields, especially the `entity_name`, to overturn this assumption and declare a match.

4.  **Format the Output**: **You must provide your answer in a valid JSON format. The JSON object should contain two keys:**
    * `select_id`: An integer. The `id` of the candidate you've determined to be an exact match. If no exact match is found, this value MUST be `-1`.
    * `explanation`: A brief, one-sentence string explaining your reasoning. For a match, explain why they are the same entity. For no match, explain the key difference.
---
-Output Schema & Format-
Your response MUST be a single, valid JSON object that adheres to the following schema. Do not include any other text, explanation, or markdown formatting like ```json.

```json
{{
  "select_id": "integer",
  "explanation": "string"
}}
```
---
-Example-
### Example 1: Match Found
### Example 2: No Match Found
----
-Task Execution-

Now, perform the selection task based on the following data. Remember to output only a single integer.

- Input Data -
\end{PromptBox}
\caption{The prompt for entity resolution judgement, examples are omitted due to lack of space.}
\label{fig:prompt_er}
\end{figure*}

\end{document}